\documentclass[preprint,12pt]{elsarticle}
\usepackage{amsmath}
\usepackage{url}
\usepackage{longtable}
\usepackage{lscape}

\usepackage{amssymb}

\journal{Interacting with Computers}

\begin{document}

\begin{frontmatter}



\title{The efficacy potential of cyber security advice as presented in news articles}


\author[inst1, corref]{Mark Quinlan}

\affiliation[inst1]{organization={Department of Computer Science, University of Oxford},
addressline={Wolfson Building, Parks Road}, 
city={Oxford},
postcode={OX1 3QD}, 
country={United Kingdom}}

\author[inst1]{Aaron Ceross}
\author[inst1]{Andrew Simpson}


\begin{abstract}
Cyber security advice is a broad church: it is thematically expansive, comprising expert texts, user-generated data consumed by individual users via informal learning, and much in-between.  While there is evidence that cyber security news articles play a role in disseminating cyber security advice, the nature and extent of that role are not clear.  We present a corpus of cyber security advice generated from mainstream news articles.  The work was driven by two research objectives.  The first objective was to ascertain what kind of actionable advice is being disseminated; the second was to explore ways of determining the efficacy potential of news-mediated security advice.  The results show an increase in the generation of cyber security news articles, together with increases in vocabulary complexity and reading difficulty.  We argue that these could present challenges for vulnerable users. We believe that this corpus and the accompanying analysis have the potential to inform future efforts to quantify and improve the efficacy potential of security advice dissemination.
\end{abstract}

\begin{highlights}
\item News-mediated security advice has been sharply increasing since 2018.
\item Many cyber security news articles have a high level of specificity. 
\item Subject-specific terminology within our security news articles is continuously evolving.
\item Cyber security news is often of short length and low readability, with a negative impact on efficacy potential.
\item The research indicates increasingly diversified interest in goal-specific cyber security advice.

\end{highlights}


\end{frontmatter}


\section{Introduction} \label{intro}
 
The usable security field is now nearly 30 years old~\cite{8996098}, although the underlying concepts are much older.  For example, Auguste Kerckhoffs acknowledged the role of the user in successful security implementation as early as 1883 when he published his seminal piece on military cryptography~\cite{kerckhoffs1883cryptographie}. Since then, users have become active participants in ensuring the security of systems and important mediators of experts' security advice~---~which, in the context of this paper, we define as explicit instructions intended to make recipients more secure, once implemented~\cite{Redmiles, pfleeger2014weakest, Herley}.

The Internet is awash with implicit and explicit security advice, disseminated both by experts and by other users.  Much of this advice has its limitations.  For example, while the underlying threats may be common, the advice provided to deal with these threats can differ significantly, both in terms of wording and implied level of urgency.  Explicit security advice that is either too abstract or too vague, or that assumes a high level of security knowledge or expertise, can fail to serve as \emph{usable advice}, as characterised by Kerckhoffs: ``the system must be easy to use and must neither require stress of mind nor the knowledge of a long set of rules''~\cite{kerckhoffs1883cryptographie}. 

Unfortunately, because individuals face a wide range of cyber security threats~---~including viruses~\cite{al2018cyber}, bot-nets~\cite{bertino2017botnets}, port-scanners~\cite{viet2018using}, spyware~\cite{karthick2017android,herley2015spyware}, malware~\cite{karthick2017android}, stalkerware~\cite{khoo2019installing}, and rootkits~\cite{karthick2017android,wang2016targeted}~---~it is difficult for them to develop sufficient `knowledge' of the `rules' even when advice is presented clearly and thoroughly. Furthermore, there is the question of what individual users do with the advice they receive.  They may, for example, reject advice they deem too difficult to implement~\cite{rader, Herley} or delay implementing advice because they underestimate the consequences of losing control of their data~\cite{brandimarte2013misplaced}.  These complications are exacerbated by the ongoing centralisation of services~\cite{barnes2006privacy} and heavy use of social media platforms~\cite{thakur2019cyber}, which may inspire complacency through habituation while introducing novel risks.  There is evidence that some users register threats on a subconscious level, raising the possibility that they will ignore novel threats. Such so-called `security fatigue' has been discussed by authors such as Furnell and Thomson~\cite{furnell2009recognising}.

Despite the significance of these problems, there have been relatively few quantitative studies analysing the security advice literature available to what we might term `everyday' users.  Notable examples of such studies include that of Redmiles and colleagues~\cite{Redmiles}, which explores how individual users actualise and perceive their own cyber security capabilities (which often derive from informally learnt advice), and that of Renaud and Dupuis~\cite{renaud2019cyber}, which identifies large sub-concepts and classifications in usable-security.  We endeavor to bridge these contributions by providing additional context about the current security advice environment. Our research is guided by the following research objectives (to which we return in Section~\ref{discussion}):
\begin{itemize}
    \item\textbf{RO1:} What kind of informally learnt and actionable security advice most often appears in news articles?
    \item\textbf{RO2:} What is the efficacy potential of this security advice as consumed by an individual user?
\end{itemize}

We describe the process we undertook to assemble a corpus of security advice that reflects the advice presented to individual users daily within their typical informal learning environment. The corpus spans a 24-month period and was assembled from a data-set containing 15,422 (English language) news and online magazine articles from North American (US and Canada) and UK-based sources, as well as some historical data stretching back to 2000.  We ascribed broad classifications to specific advice, ascertained the dominant methods of advice construction and dissemination, and analysed their potential efficacy potential over time. Our hope is that this corpus will help to lay the foundations for further work on quantifying and improving the efficacy potential of security advice dissemination.

The remainder of this paper is organised as follows. In Section~\ref{background} we provide the motivation for, and the background to, our contribution, and define the terms of interest.  In Section~\ref{methodology} we describe the process used to create the data-set, and our data cleansing process that develops the data-set into the corpus.  In Section~\ref{data} we present our results and perform an initial analysis, before, in Section~\ref{discussion}, discussing the results in relation to our research objectives.  Section~\ref{limits} considers limitations to the study and considers possible future directions.  We  conclude the paper in Section~\ref{conclusions}.

\section{Background and motivation}\label{background}

In this section we discuss the background to, and the motivation for, the work described in this paper.  We start (in Section~\ref{whatisadvice}) by contextualising the term \emph{security advice}. We define what security advice is and how security advice is learnt.  We then consider the individual user and attempt to define a persona to represent this user (in the broadest possible sense) with a view to capturing their media and information landscapes.

\subsection{What is security advice?}\label{whatisadvice}

The concept of \emph{advice} is difficult to define, partly because it pertains to almost every discipline involving human behaviour~\cite{ottaviani2006professional, bonaccio2006advice}.  \emph{Security advice} is particularly complicated, as it derives from diverse disciplines, including psychology~\cite{wiederhold2014role, dreibelbis2018looming}, medicine~\cite{yuan2018standards}, and computer science.\footnote{Given that cyber security is the discipline of concern in this paper, we shall refer to it simply as `security'.}  Each of these disciplines has its own approaches to security advice, and, as such, it would be foolish to attempt a comprehensive review of this topic covering all of these fields.  Rather, we make reference to those contributions that helped to frame, motivate and scope the present study. 

In this paper, rather than taking a broad view of security advice, we limit our concerns to what we term \emph{professional security advice}.  {\c{C}}elen et al.~\cite{ccelen2010experimental} describe the term \emph{professional advice} as ``advice rendered by experts'' (see Section~\ref{experts} for a consideration of experts), which is then disseminated to individual users. Professional security advice is designed to alleviate security challenges, but, because this must often occur through the mediation of individual users, these instructions are characteristically \emph{persuasive} and typically \emph{explicit}, rather than \emph{implicit}. 

Explicit advice often takes the form of a verbal or written appeal, such as \emph{Keep your browser up to date at all times}. There are even different varieties of appeals, as with \emph{Keep your browser up to date at all times}~\ldots~\emph{or else your browser will get infected}, which constitutes a \emph{fear appeal}~\cite{lawson2016cyber,bada2019cyber}. Implicit advice often takes the form of a threat message~\cite{renaud2019cyber}, such as \emph{Downloading a file from an untrusted source is risky}~\cite{fan2011online}.

Based on this distinction and the definitions prevalent in the literature~\cite{Redmiles,pfleeger2014weakest,Herley,rader,redmiles2016think,ion,das2018breaking}, we define \emph{security advice} thus: 
\begin{quote}
    A written instruction, provided by a trusted and professional source, with the explicit goal of enabling the recipient to be more secure once they execute the instruction.
\end{quote}{}

\subsection{The expert} \label{experts}

Note that this definition depends on the intent of the advice, rather than the outcome.  This is because professional security advice cannot ensure the success of its security recommendations as these are often developed based on limited observational data~\cite{abomhara2015cyber}.  For example, it is difficult to quantify threat-related data, such as the chance or probable extent of a malicious actor attack~\cite{tregear2001risk, cashell2004economic, jang2014survey, wagner2019automatic}.  Similarly, it is difficult to calculate the loss of customer confidence following such threats~\cite{smith2004cybercriminal}.   

Although the projected costs of hardware and/or software development to alleviate security threats might be well established in particular cases, their associated indirect costs and downstream issues are not~\cite{wagner2019automatic}.  Despite these uncertainties, professional security advice must appear authoritative, and this gap means that some professional advice may prove ineffective.  Furthermore, it is sometimes the case that some security artefacts are outdated or redundant almost immediately due to the ever-improving armoury tools and techniques available to malicious actors~\cite{chen2012business, wang2014network, casas2017network}.  It is unfortunately the case that this arms race or game of `whack-a-mole' is the environment within which most security advice is created and disseminated~\cite{von2013information}. 

Within the context of this paper, we believe it to be important to provide an explicit definition of an expert.  First, expertise as a concept can be divided into two distinct categories: expertise as a function of \textit{What someone knows}, and expertise as a function of \textit{What someone does}.  In the former, we are interested primarily in the expert's epistemic knowledge of a particular domain --- in this case, their capacity to provide justifications for any given range of ideas and knowledge \cite{weinstein1993expert}. 

Many organisations and public-facing institutions employ security experts that provide insight and compose security advice based on considerable expertise gained from education and/or industry experience \cite{caldwell2013plugging}.  Shortages of such security experts 
are regularly reported 
\cite{park2012analysis, miller2016modelling, vsorgo2017attributes, li2019data}, and this has led to an requirement for additional tools and techniques that can be used to aid current security experts in their work \cite{li2019data,mindermann2016easily}.  Shires~\cite{shires2020cyber} assessed the difficulty in establishing a firm definition of cyber security experts; an analysis of self-described practices within media highlighted a varied perception of how they operate in terms of acquiring and disseminating information.  Frey et al.~\cite{frey2017good} reported upon a test in which participants possessing several levels of cyber security knowledge were tested. The self-identified security experts tended to achieve poor scores in the test: ``they tended to display a strong interest in looking up advanced technological solutions rather than intelligence gathering''~\cite{frey2017good}.

Within almost any field in which professional advice is rendered, two forces of market are always at play.  The first concerns reputation, in which professionals are interested in how their advice is seen to be well-informed by colleagues and recipients~\cite{ajzen, lahlou, pfleeger}; the second concerns competition, which, within our context (as our advice is provided by experts working within media settings), relates to how competition between the advice providers can distort information before it reaches the intended recipient.  For example, professional advice can often take the shape of a contest, in which the experts are evaluated on the basis of their opinions.  (An example of this phenomenon can be found within the financial markets, where the Wall Street Journal Forecasting survey pits analysts and experts against each other and provides rankings~\cite{malmendier2007small, guan2018regulations}.)  Both of these kinds of influences have the potential to alter the contents of any advice rendered.  Given the 
indeterminacy, incompleteness or sometimes faultiness of the data used to generate expert advice,
alongside an unknown mixture of experts with a particular education and/or occupational background, we consider the security expert who renders our advice to be \textit{an individual who creates cyber security knowledge out of a mixture of epistemological and performative expertise}.  Their backgrounds and motivations are otherwise opaque to us --- as they may well be to the reader of the news articles --- and this forms a limitation on how we may perceive expertise in this field. 

Of course, individual users do not always get their advice from experts~\cite{ccelen2010experimental, reeder2017152, rader2015identifying}. Indeed, some may not receive information from professional sources at all~\cite{Redmiles, rader, redmiles2016think, redmiles2}.  Such users may rely on information sourced from their local environment, which we might characterise as \textit{na\"{i}ve advice}~\cite{schotter2003decision}. Contrary to this label, na\"{i}ve advice has certain efficiency-enhancing properties when used in negotiations~\cite{steinel2007effects}, public-good experiments~\cite{chaudhuri2011sustaining}, and certain types of games~\cite{kuang2007effective}.
 
Although not the focus of this paper, na\"{i}ve advice must be taken into account when we examine the recipients of security advice and how they respond to such advice.

\subsection{Who are the consumers of security advice?} \label{theusers}

Based on existing literature \cite{milne2009toward, howe2012psychology, nthala2017if}, we define consumers of security advice as `individual users' who own systems, devices, and/or services that maintain internet connectivity.  An individual user can be of any gender, age group, or professional background.  We include corporate users in this definition because their learning habits extend beyond their formal corporate learning environments~---~that is, they may engage in informal learning outside of work environments~\cite{garrick1998informal}. 


A user may design their home environment to facilitate actions like activity planning, online shopping, interpersonal communication or transmission of sensitive information (such as medical data)~\cite{nthala2017if}.  Each technology set-up can be extremely unique, akin to a fingerprint, making it difficult to assess the risks and vulnerabilities relevant to a particular space~\cite{byrne2012perceptions, nthala2017if}.

Given this variation in system complexity and user activity, each individual user must assume a degree of responsibility for the continued maintenance and integrity of their network~---~and, by extension, for the network overall \cite{pfleeger,shillair2015online, west}. Because networks are permeable, typical users may compromise their own security \emph{and} the security of others by unwittingly granting system access to malicious actors (for example, by downloading files without scanning them \cite{fan2011online}) or by failing to detect the presence of bot-nets in a slowly running system (which can destabilise large swathes of the overall network)~\cite{burghouwt2011towards}.

At the same time, the complexity of the technology environment and the diversity of online tasks makes it difficult for individual users to protect their online assets. This, paired with the (perceived) complexity of security precautions and the sheer variety of security advice and related decisions, leads individual users to report low confidence in their own decisions and in their capacity to secure their own domains \cite{shillair2015online,al2009threats}. 

\subsection{How do individuals consume security advice?} \label{howisitlearnt}

There are many ways in which individual users can encounter new security advice, but most involve some degree of \textit{formal} and \textit{informal learning}~\cite{rader,renaud2019cyber,haney2018s,stanton2016security}. 

Formal learning occurs through structured courses in an online or in-person classroom environment, usually followed by an assessment~\cite{hight2005importance, caballero2017security}.  For example, the delivery of security awareness programmes such as SETA (Security Education, Training and Awareness)\footnote{\url{https://livlab.org/seta/}} occurs within organisations and includes classes that train employees to recognise threats.  These training programmes tend to focus on compliance with corporate policy~\cite{lee2016understanding}, and they evoke generic situational awareness~\cite{lee2016understanding}, rather than providing specific contexts and situations from which an individual user can learn. 

Informal learning is unstructured, occurring outside of formal education contexts and without direct targeted interaction with security experts~\cite{stanton2016security}.  Nevertheless, it is the primary way in which adults learn about the world around them~\cite{malcolm2003interrelationships, ollis2011learning}.  As such, it is the main mode of learning considered in this paper.  Informal learning is usually triggered by some internal or external impetus~\cite{malcolm2003interrelationships}, and it occurs primarily when individuals choose to actively seek out new ideas and advice. 

Thus, for the purposes of this paper, we distinguish \emph{informal learning} from \emph{incidental learning}~\cite{malcolm2003interrelationships, bull2008connecting} on the basis of their differing intentionality: \emph{informal learning} requires some kind of prior impetus and concerted effort, whereas \emph{incidental learning} is often a by-product of carrying out another task~\cite{malcolm2003interrelationships}.  Despite being a conscious decision, informal learning is often conducted haphazardly and influenced either by randomised chance~\cite{malcolm2003interrelationships} or by the learning behaviours of others~\cite{bull2008connecting}. 

Many of the studies concerned with individual users' security intentions frame users' situational awareness and knowledge as necessary conditions for appropriate security decisions~\cite{renaud2019cyber,forget2016or}.  Essentially, researchers assume that individuals must know about the issue at hand before they can make a reasonable decision.  Thus, when an individual is faced with a security message about a potential threat, their decision process could proceed in one of two ways.
\begin{enumerate}
    \item 
First, if they already possess prior knowledge about the threat (and, more importantly, about how to prevent it), they will take appropriate action.  This is a \emph{threat control process}~\cite{renaud2019cyber}.
    \item 
    Second, if they do not possess prior awareness or knowledge, and therefore do not know how to neutralise the stated threat, the security message may be rejected.  The individual user may instead act to control the psychological fear generated by the message (rather than the practical threat implied by its contents)~\cite{renaud2019cyber,forget2016or}.
\end{enumerate}

Individual users may initially accept security advice, but subsequently reject it if they lack relevant coping strategies and actionable means to counteract the threat, choosing to deal with the issue in some other way~\cite{Redmiles,ion,howe2012psychology}.  Arguably, then, the \emph{efficacy potential} of security advice depends on how well suited it is to a given individual's existing frame of reference.  This poses interesting problems for security advice that is disseminated to a broad audience, as is the case with media-acquired advice.  As such, we give particular consideration to the role of the media.

\subsection{The efficacy potential of advice} \label{efficacy}

According to self-efficacy theory, individual users pass judgement on their own ability to cope with a given situation, thus developing self-efficacy beliefs for a specific domain. Based on these beliefs, individual users are able to initiate and persevere with behavioural strategies that lead to successful outcomes \cite{maddux1995self, bandura1999self} Self-efficacy in these cases tends to be a generative capability that allows individual users to organise their skill-sets and beliefs, which allows for an efficacy potential for these users~\cite{bandura1999self}. 

What this means for cyber security advice is that, to enhance the efficacy potential, researchers must enact strategies which help structure and direct the behaviour of individual users towards goal setting, and measure the progress towards this goal~\cite{saks1996proactive} and many usable-security studies have investigated this, e.g.\ \cite{Redmiles, pfleeger2014weakest, warner2012cybersecurity}. Recent work such as that of Furnell has additionally highlighted through current usable-security concepts how the the field may require a return to a first-principles approach \cite{furnell2024usable}. Furthermore, self-efficacy is closely linked to motivation, with the level of self-efficacy needing to be higher in order to correspond to the difficulty of the faced problem~\cite{bandura1999self, stumpf1987self}.  As already noted, cyber security is seen as both important and complex, yet the motivation to enhance self-efficacy is limited (as explained in \cite{Herley} and explored through psychological and cultural means in \cite{halevi2016cultural}). 

\subsection{The role of the media in security advice consumption}

There are many possible sources of security advice available to individual users engaged in informal learning~\cite{rader2015identifying}, including retailers and vendors of security software and services~\cite{rader2015identifying, stanton2016security}, online sources with varying levels of expertise and credibility~\cite{stanton2016security}, governmental organisations such as NIST (in the United States)\footnote{\url{https://www.nist.gov/}} and the National Cyber Security Centre (in the United Kingdom)\footnote{\url{https://www.ncsc.gov.uk/}}, professional media services such as the BBC and the Associated Press, and online media organisations such as Ars Technica\footnote{\url{https://arstechnica.com/}},
which often create and distribute security content. 

The media and communications field has the greatest reach of all of these sources. In 2017, Ruoti et al.~\cite{ruoti2017weighing} reported that individual users primarily learnt about threats through four primary sources: advertisements, news reports, television dramas, and movies. Subsequently, in 2018, Das et al.~\cite{das2018breaking} documented that news reports about threats (including cyber threats) were among the most-shared stories between individuals.  Resources such as news outlets are particularly important for older users, especially when assessing the severity of threats and the pertinence of advice~\cite{nicholson2019if}. Even fictional news can influence individual decisions about security~\cite{fulton2019effect}.

Additionally, the media and communications field is uniquely capable of influencing public opinion \emph{about} security advice \cite{ruoti2017weighing}.  News sources facilitate group-based consensus~\cite{lasswell1948structure} and set the agenda for what is regarded as an important topic, be it a presidential election~\cite{mccombs1972agenda} or an event such as the 2017 Wannacry threat~\cite{schirrmacher2018towards}, which may be accompanied by security advice.  Given this unique influence, we accord special importance to media-acquired advice in our research.  Indeed, we would argue that researchers in usable-security have a duty to understand current practices in national and international media communications. 


Media sources are adept at controlling both of these factors, using various strategies to prime the individual user and make them feel invested in the given topic~---~regardless of whether it truly pertains to them.  This `taste-making' function complements their primary advice-creation function.  As such, media sources act as `knowledge brokers' within informal learning contexts, facilitating the one-way delivery of information, concepts, and ideas from professional sources to individual users~\cite{wenger1998communities, meyer2010rise} and ultimately influencing the opinions, actions, and personal development of the recipients of security advice \cite{contandriopoulos2010knowledge}.  Consequently, the work described in this paper addresses both of these elements, as we endeavour to analyse both how media sources magnify the risk of certain security threats and their potentially associated mitigating strategies (through the use of an ontology to compare our results to), and how this security programming might gradually orient individual users' perception of security advice in general over time through a sentiment analysis.  

Thus, at this point, it is worth reiterating the research objectives of Section~\ref{intro}:

\begin{itemize}
    \item\textbf{RO1:} What kind of informally learnt and actionable security advice most often appears in news articles?
    \item\textbf{RO2:} What is the efficacy potential of this security advice as consumed by an individual user?
\end{itemize}

\section{Methodology}\label{methodology}

In this section we discuss the two elements that we utilised to obtain the necessary data.  The first element was a news-scraper, which was developed in Python and was designed to extract complete articles from structured data sources.  The second element was a viable search methodology, which was assembled from multiple components.  We first give consideration to the news-scraper. 

\subsection{The news-scraper}\label{scraper}

\begin{figure}[t]
\centering
\includegraphics[scale=0.3]{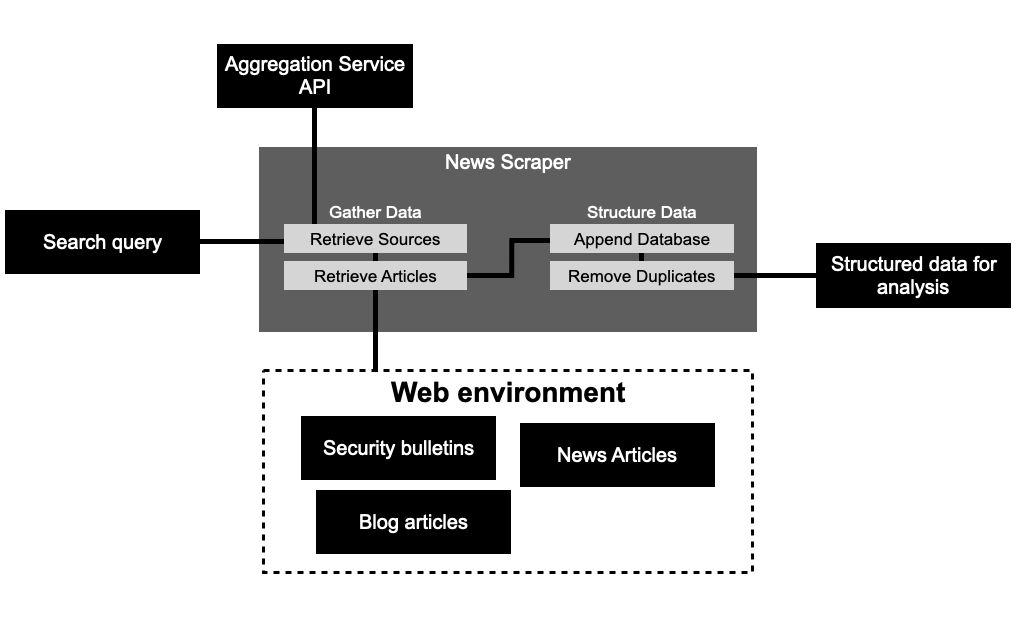}
\caption{An overview of the news-scraper tool.}
\label{Newsscraper}
\end{figure}

Web scraping is a technique that allows researchers to automate the capture of online information. Scrapers are popular tools for digital research, and they are often characterised as `outsider' tools that can be used with freely available online data~---~that is, data that does not require privileged access~\cite{marres2013scraping}. To ensure that we had enough information to answer our research objectives, we designed our news-scraper to collect as much data as possible from our news sources.  The tool's basic functional requirements are shown in Table~\ref{reqstable}; this gave rise to the abstract architecture depicted in Figure~\ref{Newsscraper}.

\begin{table}[t]
\centering
\caption{Basic functional requirements for the news-scraper.}
\begin{tabular}{ p{3cm}|p{9cm} } 
   \textbf{Requirement} & \textbf{Detail} \\
   \hline
   1 & The ability to systematically search for news articles within a set time frame utilising pre-set search queries. \\
   \hline
   2 & The ability to extract the full content from news articles. \\
   \hline
   3 & The ability to extract metadata, including publication date, author(s), titles, source names, and country of origin, for further analysis.
 \label{reqstable}
\end{tabular}
\end{table}

We utilised a news-aggregation API to filter content from a variety of unstructured and structured news sources were consistent with our definition, and we added functions to enable the complete capture of content, in accordance with Requirements 2 and 3. 

The captured data was then fed into a data-storage pipeline before being converted into a flat-file database storage solution. Incoming data was merged with existing records when required to avoid duplicate data.

\subsection{The search terms}

To fulfill Requirement 1, we followed the precedent of Schatz et al.~\cite{schatz2017towards}, who sought to derive a more precise definition of security by utilising Google Trends to automatically collect the phrases that individuals were using to search for security content.  As this had to be accomplished from the perspective of our individual user, this excluded the possibility of replicating the work of Humayun et al.~\cite{humayun2020cyber} who looked at primary studies undertaken within academia.  Instead, we followed the \textit{Systematic Mapping Study} protocol of Kosar et al.~\cite{kosar2016protocol}.

We defined a set of search and inclusion/exclusion criteria (for example, Cybersecurity OR Cyber AND Security) and additional queries containing both base search terms and queries derived from Google Trends (online OR advice OR protection OR protect OR prevent OR preventative OR tips OR email OR social network OR password OR hack OR hacked OR hacking).  

We augmented the Google Trends queries with phrases pertaining to 20 cyber security events that (1) had occurred in the previous 24 months and~(2)~had been covered by at least 10 major English-language news outlets (for the queries, please refer to Table \ref{tab:newsqueries}). Except for where it was appropriate within the event searches, all search terms were technology-agnostic --- they did not include explicit references to products or services. The news-scraper then carried out searches over a 24-month time span and returned all results that included these terms within the title or body of the content. Therefore, while not exhaustive, our corpus  represents security advice as accurately as possible within the confines of our scope.

\subsection{Cleaning the data}

First, we screened our results according to the inclusion/exclusion criteria.  These were defined as follows.

\begin{itemize}
    \item Must be a news or blog article that directly addresses at least one aspect of cyber security/contain our search terminology directly. Blog articles were limited to tutorials, editorials, tool demonstrations and discussion of technical reports.
    \item Must be written in English (due to the nature of our analysis methodology).
    \item Must be accessible, and not hidden behind a paywall or other kind of lockout mechanism (as in these cases only a few lines of text may have been retrieved). 
\end{itemize}
Any article found to be in breach of these criteria was excluded. 

In this way, we reduced the initial pool of 16,876 usable articles from our first cleaning process to 15,422 individual articles. For the remaining articles our focus and technique were informed by recent work, such as that of Satyapanich et al.~\cite{9006444}, which describes the process for extracting semantic information (such as people, places, and events) from security articles, and that of Al Moubayed et al.~\cite{Moubayed}, who used Bayesian topic modelling to ascribe classifications to, and uncover trends in, security and criminal documents.  We prepared the corpus for analysis using common data pre-processing techniques.  We utilised tokenisation to break down the text, first into sentence units and then into individual words.  We then replaced uppercase text with lowercase equivalents and removed punctuation.  We lemmatised the corpus to standardise the tense and to replace any third-person words with first-person variants.  Finally, we used a stemming technique to reduce words to their root form, where appropriate~\cite{porter2006algorithm}.

\subsection{Classifying the data}\label{ontology}


As we saw in Section~\ref{howisitlearnt}, individuals require actionable elements within their security advice. We therefore utilised an ontological framework to help us classify and integrate the data collected from the sources queried by our news-scraper.
The application of ontologies to model and reason about cyber security requirements has gained significant attention in recent years, particularly for complex systems such as critical infrastructure and smart cities. These ontologies provide a formal, machine-readable representation of key concepts and relationships, enabling precise capture and communication of security needs, automated reasoning and analysis, knowledge sharing and reuse across domains, and integration of security with other system aspects early in the development lifecycle \cite{de2021smart}. While the critical role of end-users in the overall cyber security posture of organizations and systems is increasingly recognized, current research focuses more on incorporating end-user perspectives into broader cyber security frameworks and models. For instance, some frameworks aim to identify users' security behaviors in real-time and provide targeted interventions \cite{ruighaver2007organisational}, while others leverage serious games to train users in detecting and responding to social engineering threats \cite{hendrix2016game}. However, no clear examples of ontologies solely focused on targeting end-users and their immediate requirements were found, indicating a potential gap in current research that warrants further investigation \cite{oltramari2015towards, grovs2021critical}.

In response to this we decided to perform a non-exhaustive search for an ontology which \textit{could} be applied to end-user behaviour, even if the intended target audience is not explicitly defined as such. We began by searching for ontologies using keywords such as "cyber security", "end-user", and "actionable advice", and then reviewing them against a set of selection criteria shown in Table~\ref{tab:criteria}.

\begin{table}[ht]
\centering
\begin{tabular}{|p{0.1\textwidth}|p{0.3\textwidth}|p{0.5\textwidth}|}
\hline
\textbf{ID} & \textbf{Criteria} & \textbf{Justification} \\
\hline
\textbf{1} & Evaluative in nature & The ontology should allow for an effective demonstration that a particular level of security has been achieved (efficacy potential)  \cite{souag2015security}. \\
\hline
\textbf{2} & Accessible to non-technologists & The ontology should be understandable and applicable by end-users, not just IT professionals, using clear language and unambiguous concepts \cite{kendall2019ontology}. \\
\hline
\textbf{3} & Frequently updated & The ontology should incorporate recent security concepts relevant to end-users, such as edge computing \cite{piasecki2021defence}. \\
\hline
\end{tabular}
\caption{Inclusion criteria for selecting an ontology applicable to end-user cyber security behavior.}
\label{tab:criteria}
\end{table}

In attempting to follow this criteria we found that many ontologies were indeed aimed at a technical or policy audience and often included several layers of abstraction within the work, used vaguely defined terminology or simply did not include our original requirement of actionable security advice\footnote{Examples of ontologies with many of the selection criteria but falling short of being eligible for inclusion were ENISA's \textit{IoT Security Standards Gap Analysis} (\url{https://www.enisa.europa.eu/publications/}\\\url{iot-security-standards-gap-analysis}) and a report by the UK's Department for Digital, Culture, Media and Sport, mapping security recommendations for various audiences (\url{https://www.gov.uk/government/publications/mapping-of-iot-security-}\\\url{recommendations-guidance-and-standards}).}.

\begin{figure}[t]
\centering
\includegraphics[scale=0.25]{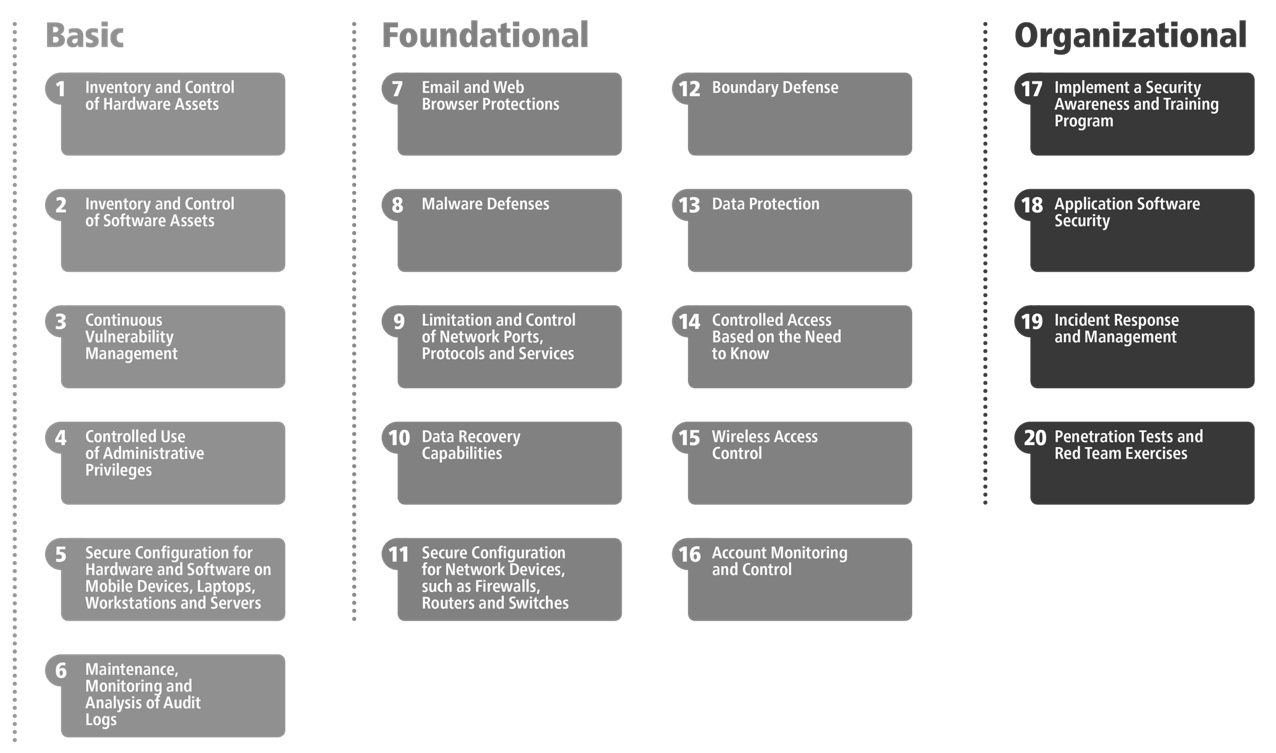}
\caption{An overview of the CIS-vectors ontological framework.}
\label{CIS-vectorsET}
\end{figure}

An ontology by the Center for Internet Security (CIS)\footnote{\url{https://learn.cisecurity.org/cis-controls-download}} was chosen, which meets all of the inclusion criteria outlined in Table~\ref{tab:criteria}. We discuss how it does so below. 

\begin{itemize}
    \item \textit{Criteria 1: }Although it was not intended specifically for individual users, this ontology prioritizes risk-based security and focuses on the practical mitigation of these risks by identifying and utilizing 20 domain-specific CIS-vectors that represent practical and actionable remedies for security threats. It does so through providing an evaluative framework that allows users to assess their security posture against specific control objectives. The controls are prioritized and provide clear guidance on essential cyber hygiene measures, making them accessible even to those with limited cybersecurity expertise. A high-level version of the framework can be seen in Figure~\ref{CIS-vectorsET}. The individual CIS-vectors are discussed in detail in Section~\ref{data}.
    \item \textit{Criteria 2:} CIS Controls provide specific, actionable guidance on the most critical steps organizations should take to tangibly improve their security, covering Criteria 2, whereas other ontologies may be more descriptive, or somewhere in between \cite{adach2022security}. As an illustrative example of Criteria 2, we refer to Woods et al.~\cite{woods2017mapping}, where the CIS ontology was shown in use with insurance underwriting professionals when selecting policy controls. 
    \item \textit{Criteria 3}: According to CIS, the CIS Controls are also frequently updated by a global community of experts to address the evolving threat landscape and incorporate recent security concepts (further confirmed by the fact that version 7.1 was utilised at the time of writing, which was then superseded by version 8) \footnote{\url{https://www.cisecurity.org/controls/cis-controls-faq}}. In addition to meeting criteria 3, the CIS ontology has been aligned with other ontological frameworks such as that of NIST to allow for easier adoption by organizations and projects\footnote{\url{https://www.cisecurity.org/blog/v7-1-introduces-implementation-groups-}\\
\url{cis-CIS-vectors/}}.
\end{itemize}

This pragmatic approach, combined with the clear tie-in to demonstrating security achievement, allows it to provide us with the requisite entity types and properties which are ascribed to individual news articles within the corpus as additional metadata. Thus, we are able to use the ontology to define the entities, relations, and other factors that can be extracted from the corpus. The ontology also allowed us to focus the corpus and to restrict our vision to the research objectives, as the language utilized within security can range from extremely specific to extremely ambiguous \cite{ruohonen2019updating}. In many cases, this range can make it difficult to apply an ontology to specific news articles within the corpus.

\subsection{Additional work to encompass null values from CIS-vectors} \label{nullvaluework}

Given our search terminology, we observed that 6,134 of the 15,422  articles (representing 36.3\% of the total corpus) contained references to any of our CIS-vectors.  We performed a second pass on the corpus, introducing additional syntactic variants of the terminology utilised within the CIS-vectors.  For example, we separated `malware defenses' into `malware AND defences', `malware defence', and `malware defense' to correct for localisation issues.  

The results of the second pass are illustrated in Figure~\ref{CISresults}.  As can be seen, the occurrence rate was subsequently 7,988 articles, or 51.7\% of the corpus.  Each of these articles contained references to one or more CIS-vectors. For the remaining 48.3\% of the corpus, we performed a Latent Dirichlet Allocation (LDA) analysis of these articles in order to generate further details, the results of which are outlined in Section~\ref{nulls}.   LDA is a statistical modeling tool that allows for the discovery of otherwise abstract topics within text files. It provides us with both a topic-per-word and topic-per-document model. To ensure the accurate selection of topic numbers and models, we followed the methodologies proposed by Cao et al.~\cite{cao2009density} and Deveaud et al.~\cite{deveaud2014accurate}. 

\section{Results}\label{data}

\begin{figure}[t]
\centering
\includegraphics[scale=0.35]{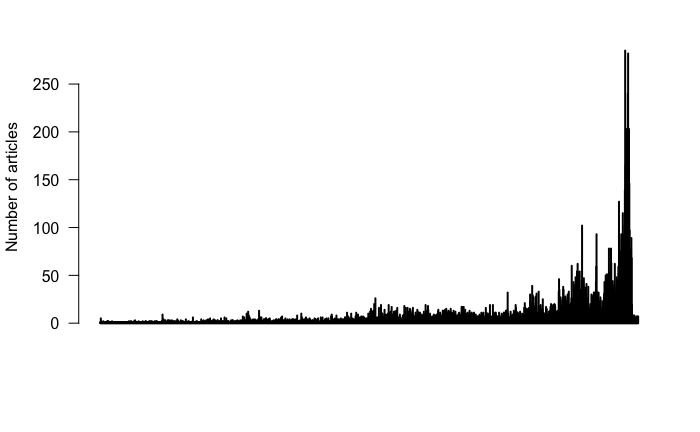}
\caption{Articles published per day between January 2015 and December 2020.}
\label{daypub}
\end{figure}

In this section we give consideration to our results. First, we corroborate the findings of Alagheband et al.~\cite{alagheband2020time}, which indicated that coverage of security topics in the New York Times has steadily increased over the last decade.  Figure~\ref{daypub} highlights this increase over time: the ``vast terra incognita of print''~\cite{taylor2004victorians}. The data also exposed the sheer diversity of publishers, ranging from traditional outlets such as the BBC news and CNN through to specialty security blogs. Even so, we must acknowledge that this list is inevitably incomplete, as our search methodology, while extensive, was non-exhaustive, and it was limited to English-language media.

Next, we identify the prevalence and features of `ideal' news articles in our corpus and use this information to help answer our research objectives. An ideal news article must contain a summary of the information that an individual user requires (in this case, regarding security advice), eliminating irrelevant and redundant information wherever possible \cite{goldstein1999summarizing}. To determine the prevalence of such articles in our corpus, we first utilised our CIS-vectors to ascertain how many of the articles contain content-specific vocabulary that users may expect to find within these articles, and we performed additional analysis on those articles that contained no such terms.  We then derived statistics pertaining to sentence length and vocabulary size, which we then compared to third-party corpora (where available).  Finally, we utilised sentiment analysis as an efficacy potential measurement tool, building on work by Kalra and Prasad~\cite{kalra2019efficacy}, who used it for stock market assessments. This was done to decipher any trends that could inform our efficacy potential research question.  

\subsection{CIS-vector occurrences} \label{CISresults}

\begin{figure}[t]
\centering
  \centering
  \includegraphics[scale=0.35]{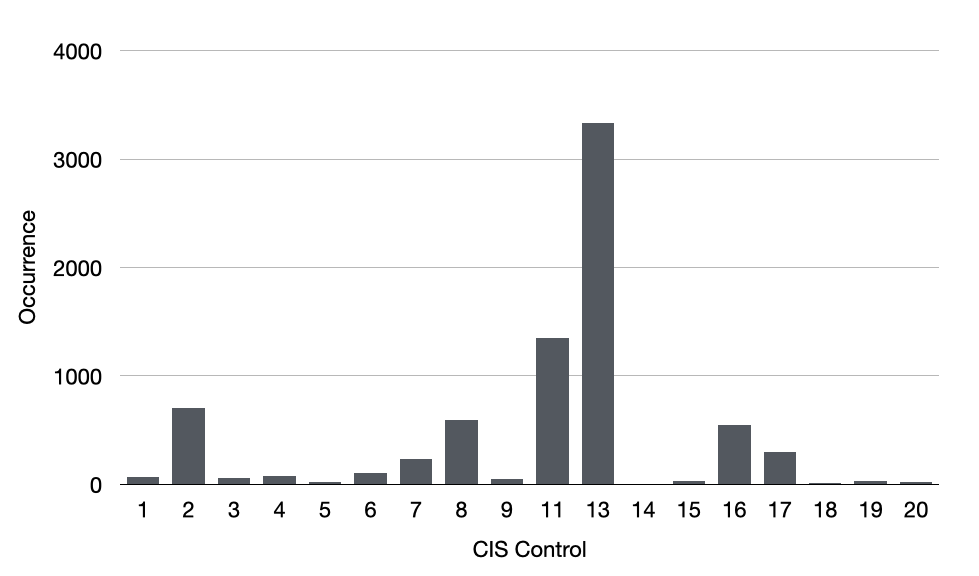}
  \label{fig:sub2}
\caption{The occurrence rate of CIS-vectors. Null values are excluded.}
\label{OCCURANCE}
\end{figure}

Figure~\ref{OCCURANCE} highlights the occurrences of our CIS-vectors in the corpus. The most-used CIS-vectors were CIS-13 (Data protection), CIS-11 (Limitation and control of network ports, protocols, and services), and CIS-2 (Inventory and control of software assets). CIS-13 highlights the growing trend towards data protection awareness and its relevance for individual users; it occurred 0.4 times per article, on average. 

Delving deeper into the reasons for this expanding data protection coverage, we find that, between 2018 and 2019, the most significant topics were related to data breaches, data protection guidelines for individuals and organisations (such as the EU's General Data Protection Regulation (GDPR)\footnote{\url{https://gdpr-info.eu/}}), and data privacy-related security advice for social media users. In 2020 there was a shift towards protecting health-related data in medical contexts, with advice and threat messaging geared towards disease contact and exposure tracing applications, such as those mentioned by Yasaka et al.~\cite{yasaka2020peer}. 

\begin{figure}[t]
\centering
  \includegraphics[scale=0.30]{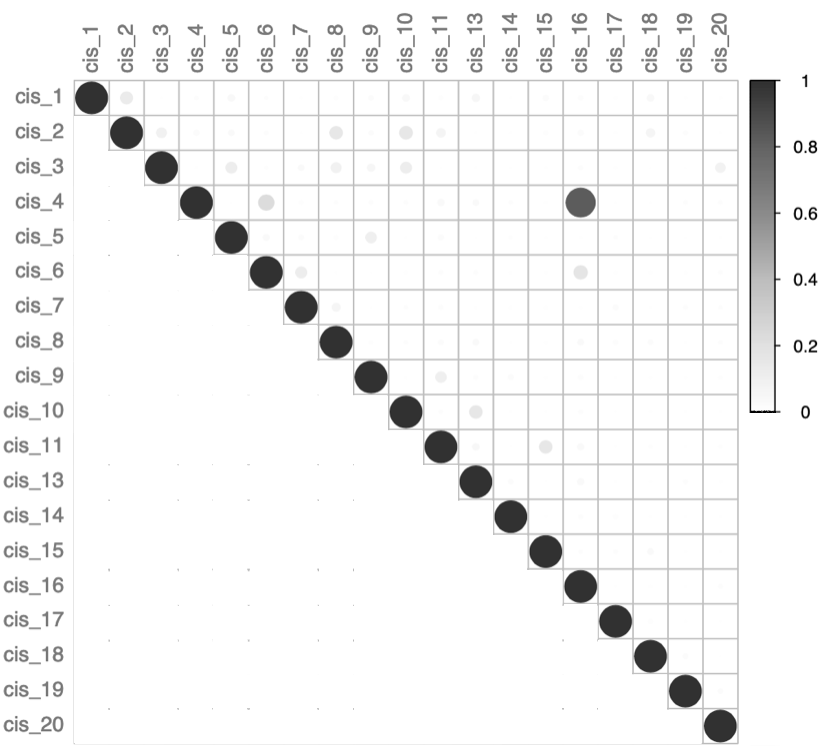}
 \caption{A correlation plot, highlighting in particular the strong correlation between CIS-4 and CIS-16.}
 \label{CORPLOT}
\end{figure}

CIS-11 indicates network security-related information and advice, and its occurrence rate increased significantly between 2019 and the end of 2020.  At least one publication (\cite{lindner2020tor}) notes a similar increase in interest.  Again, we found that most of this network security advice was related to privacy, and it appeared in texts ranging from technical articles to installation guides for the Tor Project. In many cases, these articles contained more difficult vocabulary and technical terminology than the average publication.

CIS-2 pertains to software assets and their associated CIS-vectors, and it proved to be one of the most diffuse topic. In our corpus, we found articles linked to Internet of Things home security, smart grid and connected vehicle software, and security issues that arise in connection with these devices and services. 

Correlations between the CIS-vectors are depicted in Figure~\ref{CORPLOT}.  The correlations were weak across the corpus, with one notable exception: the correlation between CIS-16 (Account monitoring and control) and CIS-4 (Controlled use of administrative privileges). Though CIS-16 appeared more frequently overall, tokens associated with both vectors appeared consistently between articles.

\subsection{Articles containing no CIS-vectors} \label{nulls}

Table~\ref{topic model} lists the most common topics that occurred in those articles that featured no CIS-vectors from our classification (representing 48.2\% of the corpus). The topics were derived through LDA topic modelling, as described in Section \ref{nullvaluework}.

\begin{table}[t]
  \centering
      \caption{The five most common topics in the non-CIS articles.}
    \begin{tabular}{|c|c|c|c|c|} \hline
Topic 3 & Topic 6 & Topic 9 & Topic 11 & Topic 14 \\ \hline
security & safety  & cyber & trump & police \\
internet & health & security & president & crime \\
system & recovery & attacks & election &  cases\\
users & covid-19 & business & russia & issue \\
data & protection & threats & u.s. & cyber \\ \hline
    \end{tabular}

    \label{topic model}
\end{table}

We can see that, despite the absence of CIS-vectors, security is still a focal point in these articles. In these cases, though, the focus is on national (cyber) security (Topic 11), cyber crime (Topic 14), business threats (Topic 9), and health and safety issues related to cyber crime and security (Topic 6). Topic 3 embodies similar concepts as CIS-vectors CIS-13 (Data protection), CIS-11 (Limitation and control of network ports, protocols, and services), and CIS-2 (Inventory and control of software assets).

\subsection{Sentence length and vocabulary size} \label{sentences}

We use sentence length, vocabulary size, and a selection of readability scores as proxies for difficulty. 

\subsubsection{Sentence length}

\begin{figure}[t]
\centering
\includegraphics[scale=0.35]{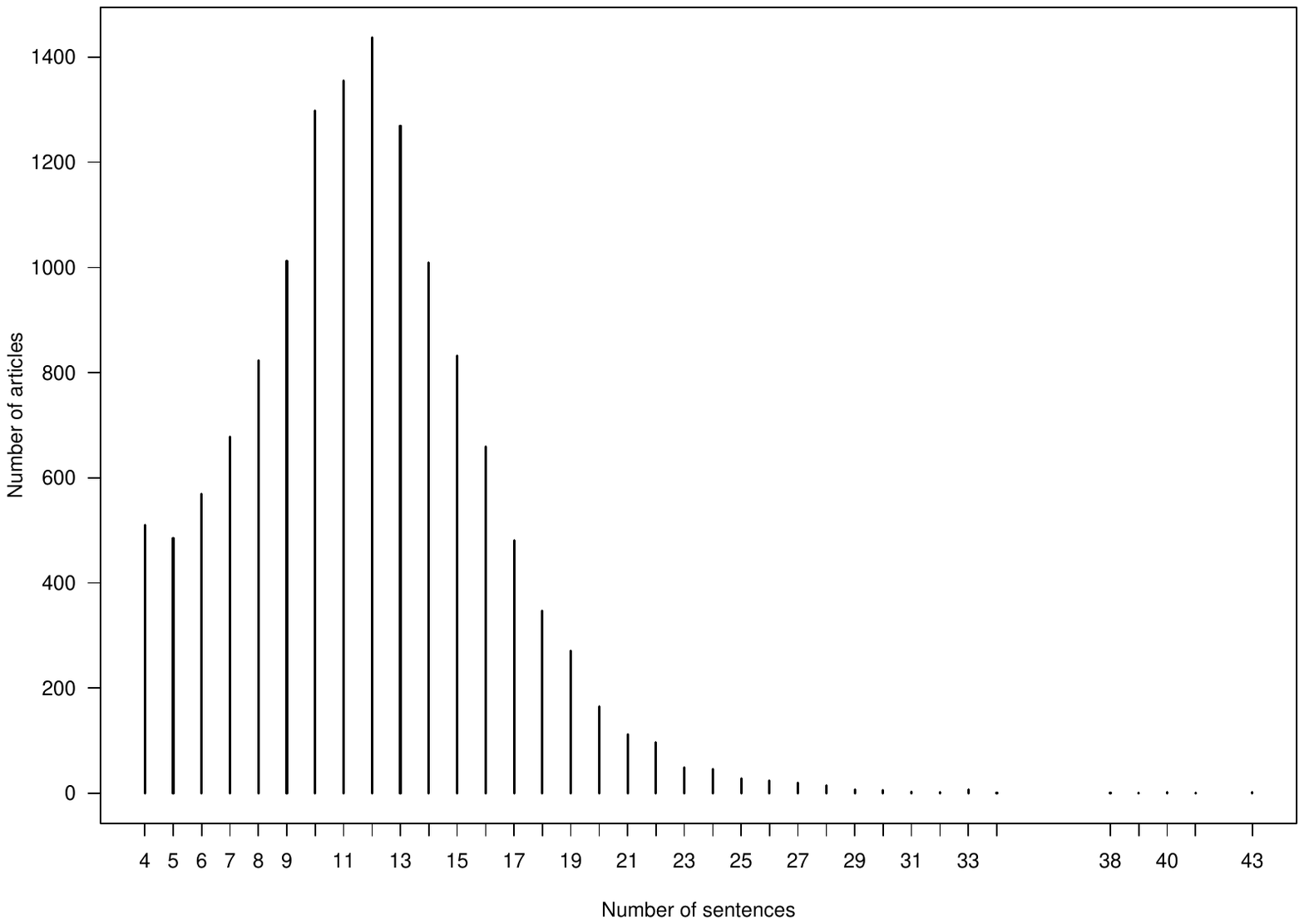}
\caption{The distribution of sentences per article.}
\label{sentence}
\end{figure}

Sentence length is an often-utilised tool in the discovery of readability within corpora~\cite{goldstein1999summarizing, lim2018understanding}. Figure~\ref{sentence} displays the average article length.  The mean article length was 9.92 sentences, and the median length was 10 sentences.  We can compare to the work of Goldstein et al.~\cite{goldstein1999summarizing} on the automated summarisation of news articles, which led to a corpus of 1,000 Reuters articles with a (post-summarisation) average length of 23 sentences.  We can also compare this to the work of Lim et al.~\cite{lim2018understanding}, whose smaller corpus yielded an average of 14 sentences per article. 

The gap between the publication of these comparators (1999 and 2018, respectively) may suggest an overall decline in the length of news articles.  It also suggests that our corpus of security-specific news is on the shorter side of the spectrum.  This last point is, however, caveated by the fact that a comparison with a more historical data and a wider potential variety of possible sources would be needed to further confirm this finding. 

\subsubsection{Vocabulary}

\begin{figure}[t]
\centering
\includegraphics[scale=0.27]{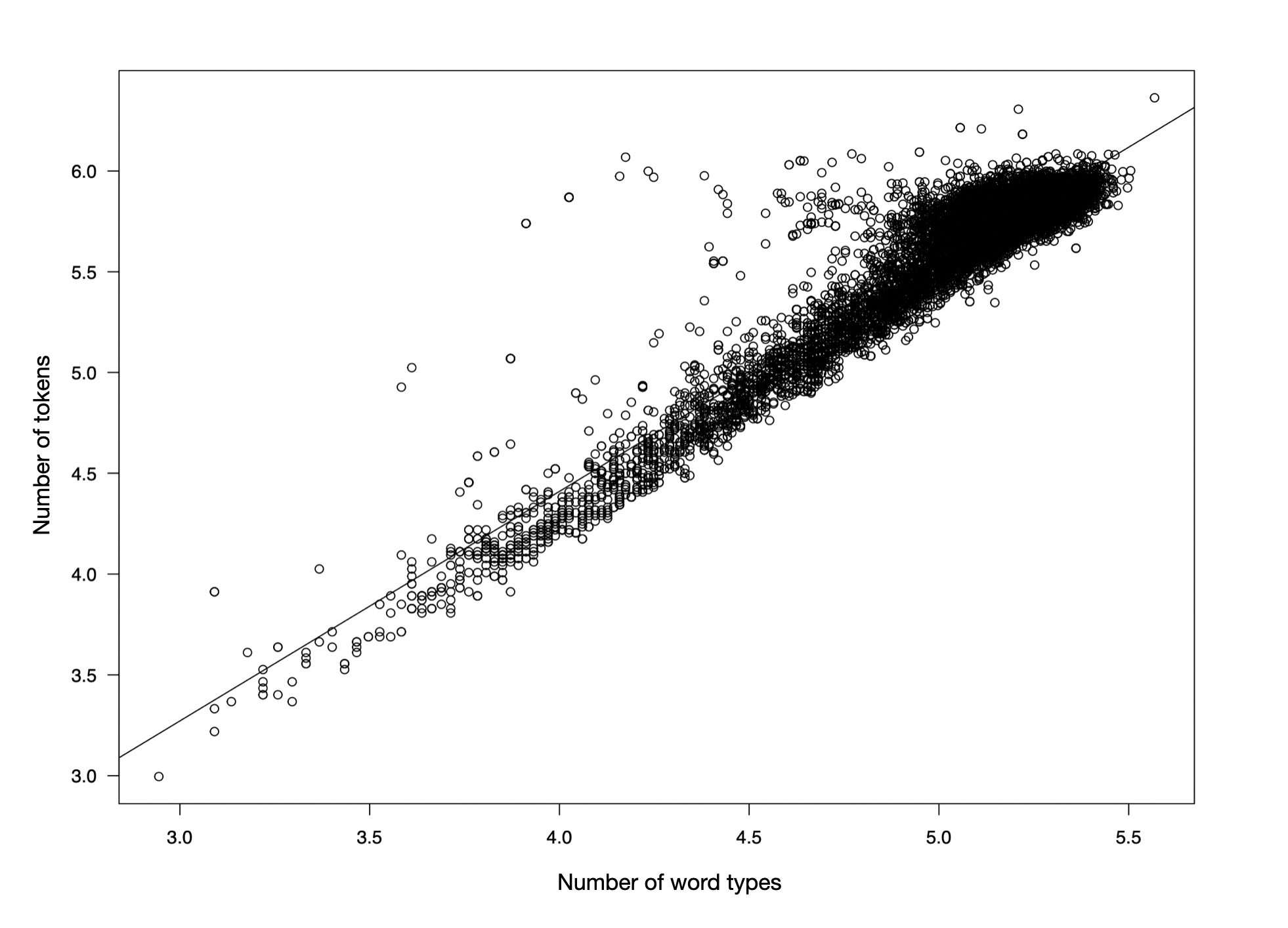}
\caption{A visualisation of Heaps' law.}
\label{HEAPS}
\end{figure}

We estimated the vocabulary growth of the corpus using Heaps' law~\cite{heaps1978information}, which describes the relationship between tokens and types. This law states that a vocabulary, expressed as \(v\) unique word types, is proportional to the power law of \(n\), the number of tokens in an arbitrary text. The relation is expressed as
\[
v = Kn^\beta
\]
Here, \(K\) is a positive constant and \(\beta\) lies between 0 and 1. In effect, as a body of text increases, the potential to discover new distinct word types decreases. In our corpus, we can see from Figure~\ref{HEAPS} that the vocabulary range largely adheres to the predicted value (black line). This means that new vocabulary terms are continually arising in the data, which could complicate users' informal learning.

\subsection{Readability scores}\label{readscore}

A readability index, such as the ones shown in Table~\ref{readabilitytable}, is an estimation of how difficult a text is to read. In online environments, it is often measured to assess click-through rates and user satisfaction~\cite{kanungo2009predicting}. Grinberg~\cite{grinberg2018identifying} utilised it, alongside sentence length, to model user engagement with news articles. As such, it is an interesting variable to consider when assessing the efficacy potential of the texts in our corpus.

\begin{table}[t]
\centering
\caption{Total Readability Metric Scores and Quartiles.}
\begin{tabular}{|l|r|r|r|r|r|}
\hline
\textbf{Metric} & \textbf{Score} & \textbf{Q1} & \textbf{Median} & \textbf{Q3} & \textbf{IQR} \\
\hline
Flesch--Kincaid & 12.52 & 10.50 & 12.85 & 15.16 & 4.66 \\
Gunning--Fog Index & 16.03 & 13.72 & 16.40 & 19.06 & 5.34 \\
Coleman--Liau Index & 13.23 & 11.65 & 13.37 & 15.01 & 3.37 \\
SMOG & 14.55 & 12.66 & 14.55 & 16.46 & 3.80 \\
Automated Readability Index & 13.00 & 11.00 & 13.00 & 16.00 & 5.00 \\
Average Grade Level & 13.87 & 11.96 & 14.18 & 16.32 & 4.36 \\
\hline
\end{tabular}
\label{readabilitytable}
\end{table}




Readability is determined by measuring a text's complexity, which is approximated via quantifiable attributes such as word length, sentence length, syllable count, and so on. The Flesch--Kincaid test~\cite{flesch2007flesch} is one of the most utilised readability tests, and it calculates readability by (1) dividing the number of utilised words by the number of sentences and (2) dividing the average number of syllables per word by the number of utilised words.  The scoring range starts at 100 for the easiest to read and descends to 0 for unreadable texts.  As an example, the combined \textit{Harry Potter} novels have a score of 72.83.  Other frequently used systems include the Gunning--Fog index~\cite{roberts1994effects}, which looks at sentence length and number of polysyllabic words; the Coleman--Liau index~\cite{coleman1975computer}, which does not assess syllables; the Automated Readability index~\cite{senter1967automated}; and the Simple Measure of Gobbledygook (or SMOG)~\cite{mc1969smog}, which utilises a similar methodology as the Flesch--Kincaid, but from sections within the text.  All of these metrics utilise a 100--0 scoring system and are broadly comparable with each another.  As such, we employ all of them in this study.

Table~\ref{readabilitytable} highlights a selection of readability scores, all utilising the same 100--0 scoring scale. To ensure the accuracy and reliability of our analysis, outliers in the readability scores data-set were identified and removed. The outlier detection was performed using the Interquartile Range (IQR) method. We calculated the first quartile (Q1) and the third quartile (Q3) for each readability measure. Outliers were defined as scores falling outside the range given by:
    \[
    \text{Lower Bound} = Q1 - 1.5 \times \text{IQR}
    \]
    \[
    \text{Upper Bound} = Q3 + 1.5 \times \text{IQR}
    \]
    where \(\text{IQR} = Q3 - Q1\).
Outliers were removed to improve the accuracy and interpretability of the results. The presence of extreme values in the data can distort statistical measures such as means and variances, and can affect the distribution and visualisation of the readability scores. After removing outliers, we re-evaluated the distribution and summary statistics of the readability scores.

\begin{table}[h]
\centering
\caption{Outlier Statistics for Readability Scores}
\begin{tabular}{|p{4cm}|r|r|r|r|}
\hline
\textbf{Measure} & \textbf{Total Outliers} & \textbf{Total Points} & \textbf{Percentage} \\
\hline
Flesch-Kincaid & 510 & 16,852 & 3.03\% \\
Gunning-Fog Index & 530 & 16,859 & 3.14\% \\
Coleman-Liau Index & 682 & 16,847 & 4.05\% \\
SMOG & 461 & 16,876 & 2.73\% \\
Automated Readability Index & 629 & 16,749 & 3.76\% \\
Average Grade Level & 531 & 16,869 & 3.15\% \\
\hline
\end{tabular}
\label{outlier_statistics}
\end{table}

\begin{figure}[t]
\centering
\includegraphics[scale=0.73]{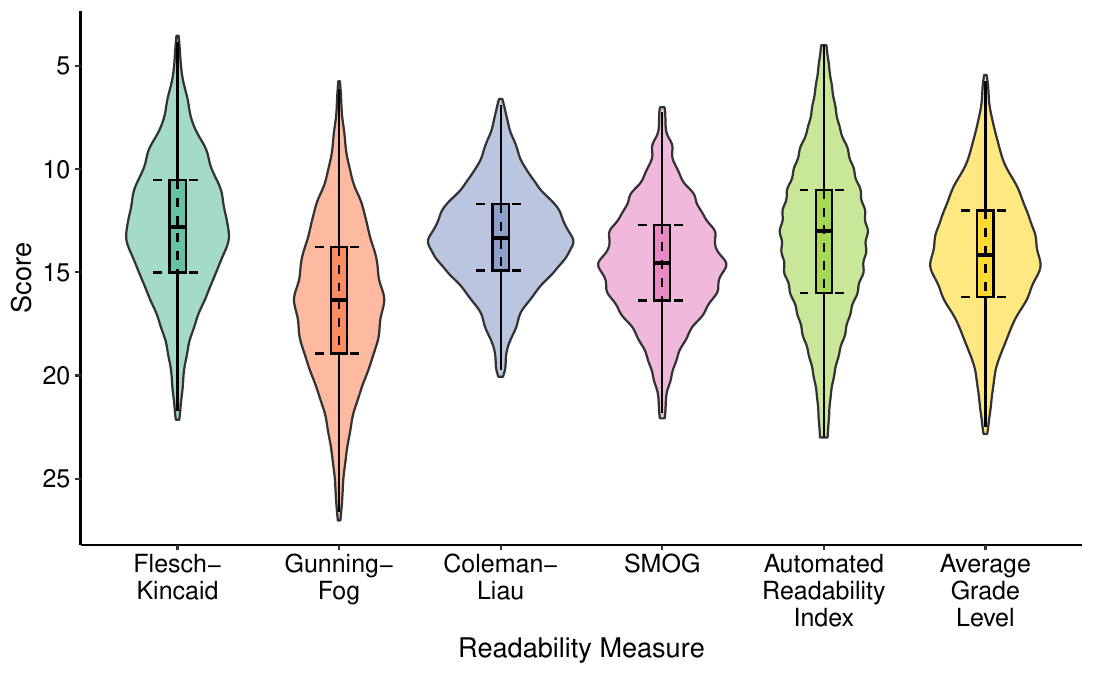}
\caption{The distributions of our readability metrics.}
\label{distro}
\end{figure}

\noindent Table~\ref{outlier_statistics} above summarises the number of outliers, total data points, and outlier percentages for each readability measure. The percentages of outliers were relatively low, ranging from 2.73\% to 4.05\%, which is generally considered manageable. Figure~\ref{distro} illustrates the distribution of readability scores after removing outliers.

The readability analysis of the corpus reveals that the text exhibits substantial complexity, as indicated by various readability metrics. The Flesch-Kincaid test'sGrade Level, with a score of 12.52, suggests that the text is suitable for readers who have completed secondary education. The Gunning Fog Index, at 16.03, implies that the text is intended for individuals with a college-level education, reflecting its complexity through longer sentences and a higher proportion of complex vocabulary. The Coleman-Liau Index, scoring 13.23, aligns with a reading level approximately one year beyond secondary school, focusing on average word length and sentence length. The SMOG Index, which stands at 14.55, indicates that a more advanced educational background is necessary for full comprehension, generally suggesting some college education or higher. The Automated Readability Index (ARI) of 13.00 further supports this, suggesting that the text is best understood by high school graduates or college students. Finally, the Average Grade Level (AGL) of 13.87 corroborates the high complexity of the text, pointing to a university-level readership.

These indices collectively illustrate that the corpus is tailored for an educated audience, characterised by complex sentence structures and sophisticated vocabulary. If the aim of the articles is to make the information more accessible to a wider audience, it may be beneficial to simplify the language and structure. However, due to the inherently complex and multifaceted nature of cybersecurity issues, such discussions will inevitably involve challenging concepts. As a result, the average casual reader may find limited value in engaging with these articles.

\subsection{Sentiment analysis}\label{sentiments}

\begin{figure}[t]
\centering
\includegraphics[scale=0.375]{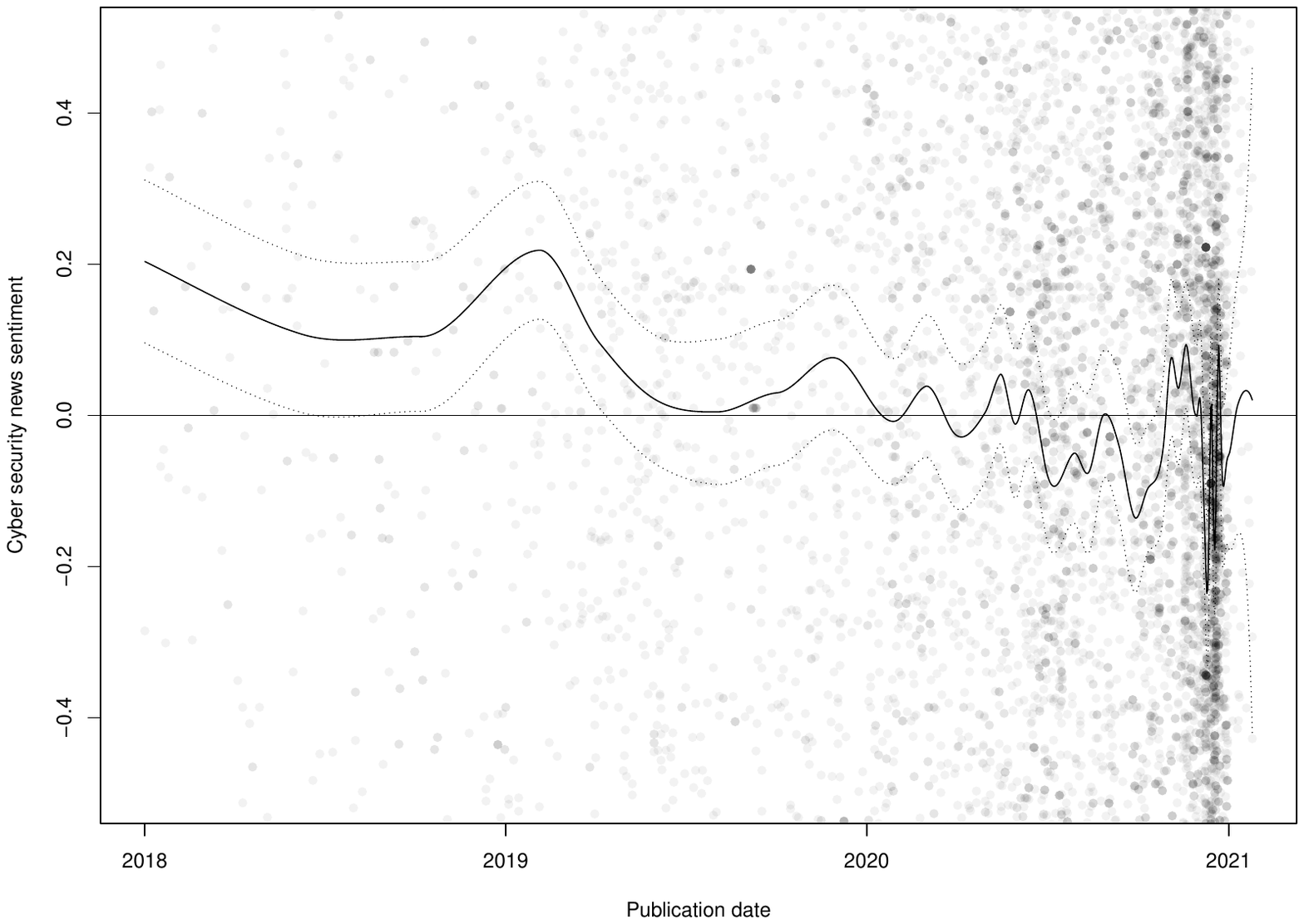}
\caption{A visualisation of changing sentiment, depicting a slight increase in negative sentiment along with a corresponding increase in article generation.}
\label{SENTIMENT}
\end{figure}

Sentiment analysis is a group of text analysis techniques that allow for the automatic derivation of sentiment (positive or negative) from large data-sets~\cite{hussein2018survey}.  Sentiment analysis is widely used across domains, from marketing~\cite{hussein2018survey} to stock market analysis~\cite{kalra2019efficacy}.  Previous sentiment analysis work within the security field has focused on predicting cyber attacks or identifying potential perpetrators, for example, by assessing sentiment in online hacker forums~\cite{macdonald2015identifying}.  The lexicons generated from these studies (for example, those pertaining to sentiment within political analysis of sovereign cyber capabilities) are of limited use within our work, as their terminology often differs substantially from what could be construed as `security advice' based on our definition.  As such, we utilised Latent Semantic Scaling (LSS), which is a semi-supervised technique for scaling documents based on work by Deerwester et al.~\cite{deerwester1990indexing}. It allows for a limited set of pre-generated seed words, which are words embedded with a specific positive or negative value. To produce our small library of seed words, we utilized the SENTPROP~\cite{hamilton2016inducing} framework. We chose this framework as it combines word-vector embeddings with a label propagation approach, which are well-known techniques to generate seed-word libraries. Additionally, SENTPROP can generate accurate results with smaller corpora.~\cite{hamilton2016inducing}. In our system, the overall sentiment of a news article is correlated with the sentiments of individual words within that article, thereby allowing for a sentiment polarity check.

The results of this sentiment analysis process can be seen in Figure~\ref{SENTIMENT}. The scores suggest an overall decrease in positive sentiment over the time period; however, these results are not statistically significant, likely because (1) the increase in published articles over the time period distorted the results and (2) a high \textit{p}-value deviated significantly across the standard \textit{alpha} value (set at 0.05).

\section{Discussion}\label{discussion}

We now consider our results in the context of the research objectives of Section~\ref{intro}. We focus on the results derived from our CIS framework and associated vectors in Section~\ref{RO1}, and ascertain how the readability, vocabulary and sentiment of the corpus affects its efficacy potential in Section~\ref{RO2}.

\subsection{What kind of informally learnt and actionable security advice most often appears in news articles?}\label{RO1}

Three overarching themes prevail in our security corpus. The first is \emph{data protection} (Theme 1), which is reflected in the strong focus on CIS-13 (Data protection) and Topic 3 of our LDA analysis. The second is \emph{physical and digital security} (Theme 2), which is supported by CIS-11 (Limitation and control of network ports, protocols, and services), CIS-2 (Inventory and control of software assets), and Topic 3 of the LDA analysis. The third is \emph{personal and collective safety} (Theme 3) in the face of personal, business, or sovereign threats to one's security, which is supported by Topics 3, 11 and~14. 

\begin{table}[t]
\centering
\caption{An overview of the themes and supporting evidence.}
 \label{themes}
\begin{tabular}{ |l|r| } \hline
   \textbf{Theme} & \textbf{Supporting evidence} \\
\hline
1.  Data protection & CIS-13, Topic 3\\
\hline
2.  Cyber-physical systems security & CIS-2, CIS-11, Topic 3\\
\hline
3.  Personal and collective safety & Topic 3, Topic 11, Topic 14 \\ \hline
\end{tabular}
\end{table}

All of these themes represent a unique set of constructs and associated user behaviours. For Themes 1 and 3, a significant driver for personal safety is privacy: ``the right of a party to maintain control over, and confidentiality of, information about itself''~\cite{oldehoeft1992foundations}. Although privacy is a significant token by itself (appearing 4,887 times in the corpus), further indirect references to it suggest that it is the underlying motivation for a significant number of data protection-related articles, be they in the realm of health data, shopping data, or, more broadly, associated with the GDPR. 

In Theme 2, personal safety entails the need for user intervention in faulty systems, either because the system cannot determine the cause of a certain threat or the appropriate corrective action to take, or, in some cases, because the system itself is acting maliciously towards the user. These articles were the most likely to contain directly actionable security advice, and thus were the most efficacious for individual users.

Theme 3 also encapsulates threats to business and sovereignty. These articles are unlikely to contain actionable security advice, but they can aid in the creation of policy~\cite{cook1998governing}, which may then lead to actionable advice. These articles may even influence public opinion regarding the (cyber) security of national sovereignty, much like how terrorism news shaped national opinion and policy, as seen in work by Gadarian~\cite{gadarian2010politics}. This cycle of influence leads to the creation of policies and legislation, such as the aforementioned GDPR, which in turn influences public awareness of potential data security threats, ultimately stimulating new forms of cyber offense and defensive capabilities.  These capabilities are then disseminated to individual users, potentially as a form of security advice. Assessing future developments within these themes and re-assessing their relevance periodically could provide a lens for evaluating the past, current, and future impact of news media on security advice efficacy potential. 

\subsection{What is the efficacy potential of this security advice as consumed by an individual user?} \label{RO2}

Many of the articles an individual user may access for cyber security advice may contain subject-specific vocabulary (such as that found within our ontological framework). Given that (1) there is limited overlap between advice sets within our ontological framework and (2) the average length of the articles in our corpus (expressed as sentence length) is shorter than the average length of comparator articles (see Section~\ref{sentences}), there appears to be a certain level of focus within the articles that could indicate efficacy potential. However, we have also seen from ontological frameworks such as the CIS-Control schema that these tools may not encompass all of the possible security vectors within the current media environment. Furthermore, these results must be qualified given our topic modelling methodology. Our application of Heaps' law highlights the growing vocabulary within our corpus, demonstrating that the subject-specific terminology in news articles on security advice is continuously evolving. This may point to an increasingly diversified interest in security advice that is tailored to a specific, predetermined goal. This encourages us to question the efficacy potential of all-encompassing frameworks such as the CIS-Control schema. 

The results of our readability tests and sentiment analysis may further challenge the efficacy potential of current media-mediated security dissemination.  We find within our corpus a trend towards high reading difficulty levels: ease of reading correlated with publication type, and news articles ranked higher on all readability indices. 
As all five of our assessment metrics reported statistically significant results with similar distribution scores (see Table \ref{readabilitytable}), we can confidently assert that just 3\% of our corpus was written at a U.S. school system 6th-grade level, which is typically the recommended reading level for standard distributed materials~\cite{kher2017readability}. Most of the articles in this corpus require a reading level of a typical college undergraduate. 

Recalling that an individual user must have (1) a sense of certainty about the content, (2) a personal interest in the content, and (3) sufficient ability to deploy the content in order to feel sufficiently compelled to act on the information, this threat control process could easily be derailed by the continued divergence and growth of subject-specific vocabulary and dense prose. Haney and Lutters~\cite{haney2018s} argue that there is a rejection threshold that informs the maintenance of security in a rapidly evolving landscape, and they maintain that individual users are approaching this threshold. 

Security is not the only specialised field that deals with these dissemination issues, and it may be helpful to observe the solutions pursued in other contexts. For example, medical advice dissemination to the general public (taken here as the equivalent of our `individual user') also involves communicating complicated concepts and extensive vocabulary to individuals who have no relevant formal training on the subject.  Britt et al.~\cite{britt2017ehealth} found that many readers stop reading medical texts if they gauge significant difficulty within the first few sentences.  Consequently, the American Medical Association (AMA) and the U.S. Department of Health and Human Services (USDHHS) have set explicit guidelines that require public-facing information to achieve a U.S.-standardised readability level of 6th grade or below~\cite{kher2017readability}.  Extrapolating these considerations to our own corpus, it would stand to reason that increasing readability to a more generally accessible level could constitute a cost-effective remedy. 

Although the overall sentiment of the corpus would not suggest that users may be being treated as an enemy (as, for example, was documented in Adams and Sasse's seminal 1999 paper~\cite{Sasse}), it does appear that what we encountered would not fulfil Kerckhoffs' criterion for ease of use. Neither would we agree that cyber security advice as portrayed in our corpus allows for self-efficacy
upon reading.  Instead, an individual user must face security topics using a multi-pronged approach, whereby self-efficacy is derived from multiple sources of increasing complexity.
If the cyber security field is to continue down the path of increased specialisation, perhaps the time has come to recognise this emerging reality and clarify --- in a transparent fashion~---~the expectations that are being placed on users.

\section{Limitations and future work} \label{limits}

The scope of this study was limited by the type and amount of information we were able to acquire to build the corpus. In our case, this meant focusing on English-language material, even though a preliminary search conducted before implementation unearthed a rich catalogue of data in other languages. This also means that our security topics, analysis, and findings likely exhibit Anglo-Saxon bias. The technical tools utilised for the readability scores were also designed for English-language articles.  There is significant scope for the enhancement of our search methodology, where for example users may only utilise the first page of any search enquiry~\cite{hochstotter2009users}.  It is our hope that this methodology be utilised to answer the same research objectives in other languages and cultural contexts.

Whilst we underscored the suitability of the CIS ontology, we also must recognise the drawbacks of this approach. The CIS ontology, although prescriptive in the manner in which it prioritises controls, lacks risk assessment specifics, and may lead to misaligned priorities and gaps as the end-user may have differing priorities. Furthermore, its suitability can also be attributed that it is due to ambiguity around its own intended target audience, and finally the CIS Controls have not undergone rigorous scientific analysis of their efficacy despite their popularity \cite{grovs2021critical}. However, as we are utilising this ontology in an effort to answer our research questions rather than appealing directly to users, and as the other ontologies we surveyed suffer from broadly similar drawbacks, we do not consider these drawbacks to be sufficient to remove it as our choice. Instead we believe that more scientific analysis and sharing of case studies on CIS Control implementations by the community would also help solidify their value proposition, and its use underscores the need for further development in user-focused cyber security ontologies which may serve as a better basis from which to base a study such as ours.

We utilised automated methodologies in order to classify topics and measure sentiment and reading difficulty, and the results are tempered by the respective limitations of these methodologies, in particular the use of a bag-of-words model, which does not capture semantic meaning or context. This method treats words as independent features, potentially leading to overestimations in mapping articles to CIS controls. For example, the mere presence of keywords might incorrectly suggest relevance to a control, disregarding nuanced meanings conveyed through context. This limitation can skew our analysis, highlighting the need for advanced techniques like word embeddings or transformer models to improve semantic understanding and mapping accuracy. Moreover, our results represent a specific snapshot in the security timeline; access to a larger historical data-set would inevitably change the overall results, potentially yielding a more statistically significant sentiment analysis.

Our approach to tackling the second research objective may limit the usefulness of our conclusions. We approximated article efficacy potential by using text analysis to predict user engagement, and we did not consider other metrics that could have enhanced the findings.  Traditionally speaking, reading-difficulty assessments in laboratory settings involve comprehension tests, eye tracking, and brain-imaging.  Knowledge of how users interact with our corpus in these terms would allow for a significantly richer analysis of security advice efficacy potential. 

The aforementioned limitations can, of course, be addressed in future research that builds upon what is presented here --- not least because our research method (described in Section~\ref{methodology}) allows for continuous data capture.  Furthermore, the data within this corpus could serve as the foundation for further analysis of security advice dissemination. Because this corpus contains a significant variety of sources, structural analysis of sentence construction for threat messaging could reveal the rhetorical structure of fear appeals, as per previous work in the field such as that of Renaud and Dupuis~\cite{renaud2019cyber}. A fear appeal is designed to motivate the reader to execute security advice, and an in-depth analysis of its features could yield results that would improve the efficacy potential of security advice dissemination. 

The corpus itself could be augmented with social media data, which would add the significant vector of digital na\"{i}ve advice~\cite{schotter2003decision}.  Bias within the articles could be used as another indicator of efficacy potential via methods like that presented by Lim et al.~\cite{lim2018understanding}.  We believe that the results of this study can provide a basis for further reflection on security advice dissemination, and that it can stimulate a conversation about individual users' learning environment.  Importantly, we hope that it serves as a point of departure for future studies.

\section{Data availability statement}

The data that support the findings of this study are available in a repository and can be accessed here: \url{https://huggingface.co/datasets/Quinm101/cybernewsarticles}.

\section{Conclusion} \label{conclusions}


We have presented work on a corpus of security advice generated from mainstream news articles as might be faced by individual users on a regular basis. The work was oriented by two questions: (1) What kind of informally learnt and actionable security advice most often appears in news articles? and (2) What is the efficacy potential of this security advice as consumed by an individual user?

We found that news-mediated security advice has been increasing since 2018, and that many such news articles focus on specific security topics. This level of focus may indicate efficacy potential. Additionally, we found that news-mediated security advice is characterised by short article length and low readability, making it difficult for many individual users to comprehend its content. We found that the subject-specific terminology within our security news articles is continuously evolving, potentially indicating increasingly diversified interest in goal-specific security advice. Again, this may increase the relative difficulty of acquiring and comprehending news-mediated security advice, with an associated impact on efficacy potential. Our approach involved using quantitative methods to yield qualitative findings. Our hope is that this research can help lay the foundations for various means of quantifying and improving the efficacy potential of security advice dissemination.





 \bibliographystyle{elsarticle-num} 
 \bibliography{cas-refs}

\begin{thebibliography}{100}
\expandafter\ifx\csname url\endcsname\relax
  \def\url#1{\texttt{#1}}\fi
\expandafter\ifx\csname urlprefix\endcsname\relax\def\urlprefix{URL }\fi
\expandafter\ifx\csname href\endcsname\relax
  \def\href#1#2{#2} \def\path#1{#1}\fi

\bibitem{8996098}
M.~Theofanos, \href{https://doi.org/10.1109/MC.2019.2954075}{Is usable security an oxymoron?}, IEEE Computer 53~(2) (2020) 71--74.
\newline\urlprefix\url{https://doi.org/10.1109/MC.2019.2954075}

\bibitem{kerckhoffs1883cryptographie}
A.~Kerckhoffs, La cryptographie militaire., Journal des Sciences Militaires IX (1883) 5--38.

\bibitem{Redmiles}
E.~M. Redmiles, S.~Kross, M.~L. Mazurek, \href{https://doi.org/10.1145/2976749.2978307}{How {I} learned to be secure: A census-representative survey of security advice sources and behavior}, in: Proceedings of the 2016 ACM SIGSAC Conference on Computer and Communications Security, CCS ’16, Association for Computing Machinery, New York, NY, USA, 2016, pp. 666--677.
\newblock \href {https://doi.org/10.1145/2976749.2978307} {\path{doi:10.1145/2976749.2978307}}.
\newline\urlprefix\url{https://doi.org/10.1145/2976749.2978307}

\bibitem{pfleeger2014weakest}
S.~L. Pfleeger, M.~A. Sasse, A.~Furnham, \href{https://doi.org/10.1515/jhsem-2014-0035}{From weakest link to security hero: Transforming staff security behavior}, Journal of Homeland Security and Emergency Management 11~(4) (2014) 489--510.
\newblock \href {https://doi.org/10.1515/jhsem-2014-0035} {\path{doi:10.1515/jhsem-2014-0035}}.
\newline\urlprefix\url{https://doi.org/10.1515/jhsem-2014-0035}

\bibitem{Herley}
C.~E. Herley, \href{https://doi.org/10.1145/1719030.1719050}{So long, and no thanks for the externalities: The rational rejection of security advice by users}, in: Proceedings of the 2009 Workshop on New Security Paradigms Workshop, NSPW ’09, Association for Computing Machinery, New York, NY, USA, 2009, pp. 133--144.
\newblock \href {https://doi.org/10.1145/1719030.1719050} {\path{doi:10.1145/1719030.1719050}}.
\newline\urlprefix\url{https://doi.org/10.1145/1719030.1719050}

\bibitem{al2018cyber}
M.~N. Al-Mhiqani, R.~Ahmad, W.~Yassin, A.~Hassan, Z.~Z. Abidin, N.~S. Ali, K.~H. Abdulkareem, \href{https://doi.org/10.14569/IJACSA.2018.090169}{Cyber-security incidents: a review cases in cyber-physical systems}, International Journal of Advanced Computer Science and Applications 9~(1) (2018) 499--508.
\newblock \href {https://doi.org/10.14569/IJACSA.2018.090169} {\path{doi:10.14569/IJACSA.2018.090169}}.
\newline\urlprefix\url{https://doi.org/10.14569/IJACSA.2018.090169}

\bibitem{bertino2017botnets}
E.~Bertino, N.~Islam, \href{https://doi.org/10.1109/MC.2017.62}{Botnets and internet of things security}, IEEE Computer 50~(2) (2017) 76--79.
\newblock \href {https://doi.org/10.1109/MC.2017.62} {\path{doi:10.1109/MC.2017.62}}.
\newline\urlprefix\url{https://doi.org/10.1109/MC.2017.62}

\bibitem{viet2018using}
H.~N. Viet, Q.~N. Van, L.~L.~T. Trang, N.~Shone, \href{https://doi.org/10.1145/3233347.3233379}{Using deep learning model for network scanning detection}, in: Proceedings of the 4th International Conference on Frontiers of Educational Technologies, 2018, pp. 117--121.
\newblock \href {https://doi.org/10.1145/3233347.3233379} {\path{doi:10.1145/3233347.3233379}}.
\newline\urlprefix\url{https://doi.org/10.1145/3233347.3233379}

\bibitem{karthick2017android}
K.~Sowndarajan, S.~Binu, \href{https://doi.org/10.1109/ICIMIA.2017.7975551}{Android security issues and solutions}, in: 2017 International Conference on Innovative Mechanisms for Industry Applications (ICIMIA), IEEE, 2017, pp. 686--689.
\newblock \href {https://doi.org/10.1109/ICIMIA.2017.7975551} {\path{doi:10.1109/ICIMIA.2017.7975551}}.
\newline\urlprefix\url{https://doi.org/10.1109/ICIMIA.2017.7975551}

\bibitem{herley2015spyware}
C.~E. Herley, B.~W. Keogh, A.~M. Hulett, A.~M. Marinescu, J.~S. Williams, S.~Nurilov, U{S} patent 9,021,590: Spyware detection mechanism (April 28 2015).

\bibitem{khoo2019installing}
C.~Khoo, K.~Robertson, R.~Deibert, Installing fear: A {C}anadian legal and policy analysis of using, developing, and selling smartphone spyware and stalkerware applications, University of Toronto Citizen Lab Report~(120) (2019).

\bibitem{wang2016targeted}
D.~Wang, Z.~Zhang, P.~Wang, J.~Yan, X.~Huang, \href{https://doi.org/10.1145/2976749.2978339}{Targeted online password guessing: An underestimated threat}, in: Proceedings of the 2016 ACM SIGSAC conference on computer and communications security, 2016, pp. 1242--1254.
\newblock \href {https://doi.org/10.1145/2976749.2978339} {\path{doi:10.1145/2976749.2978339}}.
\newline\urlprefix\url{https://doi.org/10.1145/2976749.2978339}

\bibitem{rader}
E.~Rader, R.~Wash, B.~Brooks, \href{https://doi.org/10.1145/2335356.2335364}{Stories as informal lessons about security}, in: Proceedings of the Eighth Symposium on Usable Privacy and Security, ACM, 2012, p.~6.
\newblock \href {https://doi.org/10.1145/2335356.2335364} {\path{doi:10.1145/2335356.2335364}}.
\newline\urlprefix\url{https://doi.org/10.1145/2335356.2335364}

\bibitem{brandimarte2013misplaced}
L.~Brandimarte, A.~Acquisti, G.~Loewenstein, \href{https://doi.org/10.1177/1948550612455931}{Misplaced confidences: {P}rivacy and the control paradox}, Social Psychological and Personality Science 4~(3) (2013) 340--347.
\newblock \href {https://doi.org/10.1177/1948550612455931} {\path{doi:10.1177/1948550612455931}}.
\newline\urlprefix\url{https://doi.org/10.1177/1948550612455931}

\bibitem{barnes2006privacy}
S.~B. Barnes, \href{https://doi.org/10.5210/fm.v11i9.1394}{A privacy paradox: {Social networking in the United States}}, First Monday 11~(9) (2006).
\newblock \href {https://doi.org/10.5210/fm.v11i9.1394} {\path{doi:10.5210/fm.v11i9.1394}}.
\newline\urlprefix\url{https://doi.org/10.5210/fm.v11i9.1394}

\bibitem{thakur2019cyber}
K.~Thakur, T.~Hayajneh, J.~Tseng, \href{https://doi.org/10.1109/MITP.2018.2881373}{Cyber-security in social media: {C}hallenges and the way forward}, IT Professional 21~(2) (2019) 41--49.
\newline\urlprefix\url{https://doi.org/10.1109/MITP.2018.2881373}

\bibitem{furnell2009recognising}
S.~Furnell, K.-L. Thomson, \href{https://doi.org/10.1016/S1361-3723(09)70139-3}{Recognising and addressing ‘security fatigue’}, Computer Fraud \& Security 2009~(11) (2009) 7--11.
\newblock \href {https://doi.org/10.1016/S1361-3723(09)70139-3} {\path{doi:10.1016/S1361-3723(09)70139-3}}.
\newline\urlprefix\url{https://doi.org/10.1016/S1361-3723(09)70139-3}

\bibitem{renaud2019cyber}
K.~Renaud, M.~Dupuis, \href{https://doi.org/10.1145/3368860.3368864}{Cyber-security fear appeals: Unexpectedly complicated}, in: Proceedings of the New Security Paradigms Workshop, 2019, pp. 42--56.
\newblock \href {https://doi.org/10.1145/3368860.3368864} {\path{doi:10.1145/3368860.3368864}}.
\newline\urlprefix\url{https://doi.org/10.1145/3368860.3368864}

\bibitem{ottaviani2006professional}
M.~Ottaviani, P.~N. S{\o}rensen, Professional advice, Journal of Economic Theory 126~(1) (2006) 120--142.

\bibitem{bonaccio2006advice}
S.~Bonaccio, R.~S. Dalal, \href{https://doi.org/10.1016/j.obhdp.2006.07.001}{Advice taking and decision-making: An integrative literature review, and implications for the organizational sciences}, Organizational Behavior and Human Decision Processes 101~(2) (2006) 127--151.
\newblock \href {https://doi.org/10.1016/j.obhdp.2006.07.001} {\path{doi:10.1016/j.obhdp.2006.07.001}}.
\newline\urlprefix\url{https://doi.org/10.1016/j.obhdp.2006.07.001}

\bibitem{wiederhold2014role}
B.~K. Wiederhold, \href{https://doi.org/10.1089/cyber.2014.1502}{The role of psychology in enhancing cyber-security}, Cyberpsychology, Behavior, and Social Networking 17~(3) (2014) 131--132.
\newblock \href {https://doi.org/10.1089/cyber.2014.1502} {\path{doi:10.1089/cyber.2014.1502}}.
\newline\urlprefix\url{https://doi.org/10.1089/cyber.2014.1502}

\bibitem{dreibelbis2018looming}
R.~C. Dreibelbis, J.~Martin, M.~D. Coovert, D.~W. Dorsey, \href{https://doi.org/10.1017/iop.2018.3}{The looming cyber-security crisis and what it means for the practice of industrial and organizational psychology}, Industrial and Organizational Psychology 11~(2) (2018) 346--365.
\newblock \href {https://doi.org/10.1017/iop.2018.3} {\path{doi:10.1017/iop.2018.3}}.
\newline\urlprefix\url{https://doi.org/10.1017/iop.2018.3}

\bibitem{yuan2018standards}
S.~Yuan, A.~Fernando, D.~C. Klonoff, \href{https://doi.org/10.1177/1932296818763634}{Standards for medical device cyber-security in 2018} (2018).
\newblock \href {https://doi.org/10.1177/1932296818763634} {\path{doi:10.1177/1932296818763634}}.
\newline\urlprefix\url{https://doi.org/10.1177/1932296818763634}

\bibitem{ccelen2010experimental}
B.~{\c{C}}elen, S.~Kariv, A.~Schotter, \href{https://doi.org/10.1287/mnsc.1100.1228}{An experimental test of advice and social learning}, Management Science 56~(10) (2010) 1687--1701.
\newblock \href {https://doi.org/10.1287/mnsc.1100.1228} {\path{doi:10.1287/mnsc.1100.1228}}.
\newline\urlprefix\url{https://doi.org/10.1287/mnsc.1100.1228}

\bibitem{lawson2016cyber}
S.~T. Lawson, S.~K. Yeo, H.~Yu, E.~Greene, The cyber-doom effect: The impact of fear appeals in the us cyber-security debate, in: 2016 8th International Conference on Cyber Conflict (CyCon), IEEE, 2016, pp. 65--80.

\bibitem{bada2019cyber}
M.~Bada, A.~M. Sasse, J.~R.~C. Nurse, Cyber-security awareness campaigns: Why do they fail to change behaviour?, arXiv preprint arXiv:1901.02672 (2019).

\bibitem{fan2011online}
W.~Fan, K.~H. Yeung, \href{https://doi.org/10.1016/j.physa.2010.09.034}{Online social networks—paradise of computer viruses}, Physica A: Statistical Mechanics and its Applications 390~(2) (2011) 189--197.
\newblock \href {https://doi.org/10.1016/j.physa.2010.09.034} {\path{doi:10.1016/j.physa.2010.09.034}}.
\newline\urlprefix\url{https://doi.org/10.1016/j.physa.2010.09.034}

\bibitem{redmiles2016think}
E.~M. Redmiles, A.~R. Malone, M.~L. Mazurek, I think they're trying to tell me something: Advice sources and selection for digital security, in: 2016 IEEE Symposium on Security and Privacy (SP), IEEE, 2016, pp. 272--288.

\bibitem{ion}
I.~Ion, R.~Reeder, S.~Consolvo, “... no one can hack my mind”: Comparing expert and non-expert security practices, in: Eleventh Symposium On Usable Privacy and Security (SOUPS), 2015, pp. 327--346.

\bibitem{das2018breaking}
S.~Das, J.~Lo, L.~Dabbish, J.~I. Hong, \href{https://doi.org/10.1145/3173574.3173575}{Breaking! {A} typology of security and privacy news and how it's shared}, in: Proceedings of the 2018 CHI Conference on Human Factors in Computing Systems, 2018, pp. 1--12, paper No. 1.
\newblock \href {https://doi.org/10.1145/3173574.3173575} {\path{doi:10.1145/3173574.3173575}}.
\newline\urlprefix\url{https://doi.org/10.1145/3173574.3173575}

\bibitem{abomhara2015cyber}
M.~Abomhara, G.~Køien, \href{https://doi.org/10.13052/jcsm2245-1439.414}{Cyber-security and the internet of things: vulnerabilities, threats, intruders and attacks}, Journal of Cyber-Security and Mobility 4~(1) (2015) 65--88.
\newblock \href {https://doi.org/10.13052/jcsm2245-1439.414} {\path{doi:10.13052/jcsm2245-1439.414}}.
\newline\urlprefix\url{https://doi.org/10.13052/jcsm2245-1439.414}

\bibitem{tregear2001risk}
J.~Tregear, \href{https://doi.org/10.1016/S1363-4127(01)00304-1}{Risk assessment}, Information Security Technical Report 3~(6) (2001) 19--27.
\newblock \href {https://doi.org/10.1016/S1363-4127(01)00304-1} {\path{doi:10.1016/S1363-4127(01)00304-1}}.
\newline\urlprefix\url{https://doi.org/10.1016/S1363-4127(01)00304-1}

\bibitem{cashell2004economic}
B.~Cashell, W.~D. Jackson, M.~Jickling, B.~Webel, The economic impact of cyber-attacks, Congressional Research Service Documents, CRS RL32331 (Washington DC) (2004) 2.

\bibitem{jang2014survey}
J.~Jang-Jaccard, S.~Nepal, \href{https://doi.org/10.1016/j.jcss.2014.02.005}{A survey of emerging threats in cyber-security}, Journal of Computer and System Sciences 80~(5) (2014) 973--993.
\newblock \href {https://doi.org/10.1016/j.jcss.2014.02.005} {\path{doi:10.1016/j.jcss.2014.02.005}}.
\newline\urlprefix\url{https://doi.org/10.1016/j.jcss.2014.02.005}

\bibitem{wagner2019automatic}
N.~Wagner, C.~{\c{S}}. {\c{S}}ahin, J.~Pena, W.~W. Streilein, Automatic generation of cyber architectures optimized for security, cost, and mission performance: A nature-inspired approach, in: Advances in Nature-Inspired Computing and Applications, Springer, 2019, pp. 1--25.

\bibitem{smith2004cybercriminal}
A.~D. Smith, \href{https://doi.org/10.1108/14684520410543670}{Cybercriminal impacts on online business and consumer confidence}, Online Information Review 28~(3) (2004) 224--234.
\newblock \href {https://doi.org/10.1108/14684520410543670} {\path{doi:10.1108/14684520410543670}}.
\newline\urlprefix\url{https://doi.org/10.1108/14684520410543670}

\bibitem{chen2012business}
H.~Chen, R.~H.~L. Chiang, V.~C. Storey, \href{https://doi.org/10.2307/41703503}{Business intelligence and analytics: From big data to big impact}, MIS Quarterly (2012) 1165--1188.
\newline\urlprefix\url{https://doi.org/10.2307/41703503}

\bibitem{wang2014network}
Y.~Wang, Y.~Wang, J.~Liu, Z.~Huang, \href{https://doi.org/10.1109/3PGCIC.2014.41}{A network gene-based framework for detecting advanced persistent threats}, in: 2014 Ninth International Conference on P2P, Parallel, Grid, Cloud and Internet Computing, IEEE, 2014, pp. 97--102.
\newblock \href {https://doi.org/10.1109/3PGCIC.2014.41} {\path{doi:10.1109/3PGCIC.2014.41}}.
\newline\urlprefix\url{https://doi.org/10.1109/3PGCIC.2014.41}

\bibitem{casas2017network}
P.~Casas, F.~Soro, J.~Vanerio, G.~Settanni, A.~D'Alconzo, \href{https://doi.org/10.1109/CloudNet.2017.8071525}{Network security and anomaly detection with big-dama, a big data analytics framework}, in: 2017 IEEE 6th International Conference on Cloud Networking (CloudNet), IEEE, 2017, pp. 1--7.
\newblock \href {https://doi.org/10.1109/CloudNet.2017.8071525} {\path{doi:10.1109/CloudNet.2017.8071525}}.
\newline\urlprefix\url{https://doi.org/10.1109/CloudNet.2017.8071525}

\bibitem{von2013information}
R.~von Solms, J.~van Niekerk, \href{https://doi.org/10.1016/j.cose.2013.04.004}{From information security to cyber-security}, Computers \& Security 38 (2013) 97--102.
\newblock \href {https://doi.org/10.1016/j.cose.2013.04.004} {\path{doi:10.1016/j.cose.2013.04.004}}.
\newline\urlprefix\url{https://doi.org/10.1016/j.cose.2013.04.004}

\bibitem{weinstein1993expert}
B.~D. Weinstein, What is an expert?, Theoretical medicine 14~(1) (1993) 57--73.

\bibitem{caldwell2013plugging}
T.~Caldwell, Plugging the cyber-security skills gap, Computer Fraud \& Security 2013~(7) (2013) 5--10.

\bibitem{park2012analysis}
J.-y. Park, An analysis on training curriculum for educating information security experts, Management \& Information Systems Review 31~(1) (2012) 149--165.

\bibitem{miller2016modelling}
S.~Miller, C.~Wagner, U.~Aickelin, J.~M. Garibaldi, Modelling cyber-security experts' decision making processes using aggregation operators, computers \& security 62 (2016) 229--245.

\bibitem{vsorgo2017attributes}
A.~{\v{S}}orgo, T.~Bartol, D.~Dolni{\v{c}}ar, B.~Boh~Podgornik, Attributes of digital natives as predictors of information literacy in higher education, British Journal of Educational Technology 48~(3) (2017) 749--767.

\bibitem{li2019data}
T.~Li, G.~Convertino, R.~K. Tayi, S.~Kazerooni, What data should i protect? recommender and planning support for data security analysts, in: Proceedings of the 24th International Conference on Intelligent User Interfaces, 2019, pp. 286--297.

\bibitem{mindermann2016easily}
K.~Mindermann, Are easily usable security libraries possible and how should experts work together to create them?, in: Proceedings of the 9th international workshop on cooperative and human aspects of software engineering, 2016, pp. 62--63.

\bibitem{shires2020cyber}
J.~Shires, Cyber-noir: Cyber-security and popular culture, Contemporary Security Policy 41~(1) (2020) 82--107.

\bibitem{frey2017good}
S.~Frey, A.~Rashid, P.~Anthonysamy, M.~Pinto-Albuquerque, S.~A. Naqvi, The good, the bad and the ugly: a study of security decisions in a cyber-physical systems game, IEEE Transactions on Software Engineering 45~(5) (2017) 521--536.

\bibitem{ajzen}
I.~Ajzen, From intentions to actions: {A} theory of planned behavior, in: Action control, Springer, 1985, pp. 11--39.

\bibitem{lahlou}
S.~Lahlou, M.~Langheinrich, C.~R{\"o}cker, Privacy and trust issues with invisible computers, Communications of the ACM 48~(3) (2005) 59--60.

\bibitem{pfleeger}
S.~L. Pfleeger, D.~D. Caputo, Leveraging behavioral science to mitigate cyber-security risk, Computers \& Security 31~(4) (2012) 597--611.

\bibitem{malmendier2007small}
U.~Malmendier, D.~Shanthikumar, Are small investors naive about incentives?, Journal of Financial Economics 85~(2) (2007) 457--489.

\bibitem{guan2018regulations}
Y.~Guan, C.~Li, H.~Lu, M.~Wong, Regulations and brain drain: Evidence from wall street star analysts’ career choices, Management Science, Forthcoming (2018).

\bibitem{reeder2017152}
R.~W. Reeder, I.~Ion, S.~Consolvo, 152 simple steps to stay safe online: Security advice for non-tech-savvy users, IEEE Security \& Privacy 15~(5) (2017) 55--64.

\bibitem{rader2015identifying}
E.~Rader, R.~Wash, \href{https://doi.org/10.1093/cybsec/tyv008}{Identifying patterns in informal sources of security information}, Journal of Cyber-Security 1~(1) (2015) 121--144.
\newblock \href {https://doi.org/10.1093/cybsec/tyv008} {\path{doi:10.1093/cybsec/tyv008}}.
\newline\urlprefix\url{https://doi.org/10.1093/cybsec/tyv008}

\bibitem{redmiles2}
E.~M. Redmiles, S.~Kross, M.~L. Mazurek, \href{https://doi.org/10.1109/SP.2019.00014}{How well do my results generalize? comparing security and privacy survey results from mturk, web, and telephone samples}, in: 2019 IEEE Symposium on Security and Privacy (SP), Vol.~00, IEEE, 2019, pp. 227--244.
\newblock \href {https://doi.org/10.1109/SP.2019.00014} {\path{doi:10.1109/SP.2019.00014}}.
\newline\urlprefix\url{https://doi.org/10.1109/SP.2019.00014}

\bibitem{schotter2003decision}
A.~Schotter, Decision making with naive advice, American Economic Review 93~(2) (2003) 196--201.

\bibitem{steinel2007effects}
W.~Steinel, A.~E. Abele, C.~K.~W. De~Dreu, \href{https://doi.org/10.1177/1368430207081541}{Effects of experience and advice on process and performance in negotiations}, Group Processes \& Intergroup Relations 10~(4) (2007) 533--550.
\newblock \href {https://doi.org/10.1177/1368430207081541} {\path{doi:10.1177/1368430207081541}}.
\newline\urlprefix\url{https://doi.org/10.1177/1368430207081541}

\bibitem{chaudhuri2011sustaining}
A.~Chaudhuri, \href{https://doi.org/10.1007/s10683-010-9257-1}{Sustaining cooperation in laboratory public goods experiments: a selective survey of the literature}, Experimental Economics 14~(1) (2011) 47--83.
\newblock \href {https://doi.org/10.1007/s10683-010-9257-1} {\path{doi:10.1007/s10683-010-9257-1}}.
\newline\urlprefix\url{https://doi.org/10.1007/s10683-010-9257-1}

\bibitem{kuang2007effective}
X.~J. Kuang, R.~A. Weber, J.~Dana, How effective is advice from interested parties?: An experimental test using a pure coordination game, Journal of Economic Behavior \& Organization 62~(4) (2007) 591--604.

\bibitem{milne2009toward}
G.~R. Milne, L.~I. Labrecque, C.~Cromer, \href{https://doi.org/10.1111/j.1745-6606.2009.01148.x}{Toward an understanding of the online consumer's risky behavior and protection practices}, Journal of Consumer Affairs 43~(3) (2009) 449--473.
\newblock \href {https://doi.org/10.1111/j.1745-6606.2009.01148.x} {\path{doi:10.1111/j.1745-6606.2009.01148.x}}.
\newline\urlprefix\url{https://doi.org/10.1111/j.1745-6606.2009.01148.x}

\bibitem{howe2012psychology}
A.~E. Howe, I.~Ray, M.~Roberts, M.~Urbanska, Z.~Byrne, \href{https://doi.org/10.1109/SP.2012.23}{The psychology of security for the home computer user}, in: 2012 IEEE Symposium on Security and Privacy, IEEE, 2012, pp. 209--223.
\newblock \href {https://doi.org/10.1109/SP.2012.23} {\path{doi:10.1109/SP.2012.23}}.
\newline\urlprefix\url{https://doi.org/10.1109/SP.2012.23}

\bibitem{nthala2017if}
N.~Nthala, I.~Flechais, “if it’s urgent or it is stopping me from doing something, then i might just go straight at it”: a study into home data security decisions, in: International Conference on Human Aspects of Information Security, Privacy, and Trust, Springer, 2017, pp. 123--142.

\bibitem{garrick1998informal}
J.~Garrick, Informal learning in corporate workplaces, Human Resource Development Quarterly 9~(2) (1998) 129--144.

\bibitem{byrne2012perceptions}
Z.~Byrne, J.~Weidert, J.~Liff, M.~Horvath, C.~Smith, A.~Howe, I.~Ray, Perceptions of internet threats: Behavioral intent to click again, in: Proceedings of the 27th Annual Conference of the Society for Industrial and Organizational Psychology, 2012, pp. 26--28.

\bibitem{shillair2015online}
R.~Shillair, S.~R. Cotten, H.-Y.~S. Tsai, S.~Alhabash, R.~LaRose, N.~J. Rifon, \href{https://doi.org/10.1016/j.chb.2015.01.046}{Online safety begins with you and me: Convincing internet users to protect themselves}, Computers in Human Behavior 48 (2015) 199--207.
\newblock \href {https://doi.org/10.1016/j.chb.2015.01.046} {\path{doi:10.1016/j.chb.2015.01.046}}.
\newline\urlprefix\url{https://doi.org/10.1016/j.chb.2015.01.046}

\bibitem{west}
R.~West, \href{https://doi.org/10.1145/1330311.1330320}{The psychology of security}, Communications of the ACM 51~(4) (2008) 34--40.
\newblock \href {https://doi.org/10.1145/1330311.1330320} {\path{doi:10.1145/1330311.1330320}}.
\newline\urlprefix\url{https://doi.org/10.1145/1330311.1330320}

\bibitem{burghouwt2011towards}
P.~Burghouwt, M.~Spruit, H.~Sips, Towards detection of botnet communication through social media by monitoring user activity, in: International Conference on Information Systems Security, Springer, 2011, pp. 131--143.

\bibitem{al2009threats}
A.~Al~Hasib, Threats of online social networks, International Journal of Computer Science and Network Security (IJCSNS) 9~(11) (2009) 288--293.

\bibitem{haney2018s}
J.~M. Haney, W.~G. Lutters, \href{https://www.usenix.org/conference/soups2018/presentation/haney-perceptions}{"it's scary…it's confusing…it's dull": How cyber-security advocates overcome negative perceptions of security}, in: Fourteenth Symposium on Usable Privacy and Security (SOUPS), 2018, pp. 411--425.
\newline\urlprefix\url{https://www.usenix.org/conference/soups2018/presentation/haney-perceptions}

\bibitem{stanton2016security}
B.~Stanton, M.~F. Theofanos, S.~S. Prettyman, S.~Furman, \href{https://doi.org/10.1109/MITP.2016.84}{Security fatigue}, IT Professional 18~(5) (2016) 26--32.
\newblock \href {https://doi.org/10.1109/MITP.2016.84} {\path{doi:10.1109/MITP.2016.84}}.
\newline\urlprefix\url{https://doi.org/10.1109/MITP.2016.84}

\bibitem{hight2005importance}
S.~D. Hight, The importance of a security, education, training and awareness program, november 2005, Security 27601 (2005) 1--5.

\bibitem{caballero2017security}
A.~Caballero, \href{https://doi.org/10.1016/b978-0-12-803843-7.00033-8}{Security education, training, and awareness}, in: Computer and information security handbook, Elsevier, 2017, pp. 497--505.
\newblock \href {https://doi.org/10.1016/b978-0-12-803843-7.00033-8} {\path{doi:10.1016/b978-0-12-803843-7.00033-8}}.
\newline\urlprefix\url{https://doi.org/10.1016/b978-0-12-803843-7.00033-8}

\bibitem{lee2016understanding}
C.~Lee, C.~C. Lee, S.~Kim, \href{https://doi.org/10.1016/j.cose.2016.02.004}{Understanding information security stress: Focusing on the type of information security compliance activity}, Computers \& Security 59 (2016) 60--70.
\newblock \href {https://doi.org/10.1016/j.cose.2016.02.004} {\path{doi:10.1016/j.cose.2016.02.004}}.
\newline\urlprefix\url{https://doi.org/10.1016/j.cose.2016.02.004}

\bibitem{malcolm2003interrelationships}
J.~Malcolm, P.~Hodkinson, H.~Colley, \href{https://doi.org/10.1108/13665620310504783}{The interrelationships between informal and formal learning}, Journal of Workplace Learning (2003).
\newblock \href {https://doi.org/10.1108/13665620310504783} {\path{doi:10.1108/13665620310504783}}.
\newline\urlprefix\url{https://doi.org/10.1108/13665620310504783}

\bibitem{ollis2011learning}
T.~Ollis, Learning in social action: The informal and social learning dimensions of circumstantial and lifelong activists., Australian Journal of Adult Learning 51~(2) (2011) 248--268.

\bibitem{bull2008connecting}
G.~Bull, A.~Thompson, M.~Searson, J.~Garofalo, J.~Park, C.~Young, J.~Lee, Connecting informal and formal learning experiences in the age of participatory media, Contemporary Issues in Technology and Teacher Education 8~(2) (2008) 100--107.

\bibitem{forget2016or}
A.~Forget, S.~Pearman, J.~Thomas, A.~Acquisti, N.~Christin, L.~F. Cranor, S.~Egelman, M.~Harbach, R.~Telang, Do or do not, there is no try: user engagement may not improve security outcomes, in: Twelfth Symposium on Usable Privacy and Security (SOUPS), 2016, pp. 97--111.

\bibitem{maddux1995self}
J.~E. Maddux, Self-efficacy theory, in: Self-efficacy, adaptation, and adjustment, Springer, 1995, pp. 3--33.

\bibitem{bandura1999self}
A.~Bandura, W.~H. Freeman, R.~Lightsey, Self-efficacy: The exercise of control, Springer, 1999.

\bibitem{saks1996proactive}
A.~M. Saks, B.~E. Ashforth, Proactive socialization and behavioral self-management, Journal of Vocational behavior 48~(3) (1996) 301--323.

\bibitem{warner2012cybersecurity}
M.~Warner, Cyber-security: A pre-history, Intelligence and National Security 27~(5) (2012) 781--799.

\bibitem{furnell2024usable}
S.~Furnell, Usable cybersecurity: a contradiction in terms?, Interacting with Computers (2024) iwad035.

\bibitem{stumpf1987self}
S.~A. Stumpf, A.~P. Brief, K.~Hartman, Self-efficacy expectations and coping with career-related events, Journal of Vocational Behavior 31~(1) (1987) 91--108.

\bibitem{halevi2016cultural}
T.~Halevi, N.~Memon, J.~Lewis, P.~Kumaraguru, S.~Arora, N.~Dagar, F.~Aloul, J.~Chen, Cultural and psychological factors in cyber-security, in: Proceedings of the 18th International Conference on Information Integration and Web-based Applications and Services, 2016, pp. 318--324.

\bibitem{ruoti2017weighing}
S.~Ruoti, T.~Monson, J.~Wu, D.~Zappala, K.~Seamons, \href{https://www.usenix.org/conference/soups2017/technical-sessions/presentation/ruoti}{Weighing context and trade-offs: How suburban adults selected their online security posture}, in: Proceedings of the Thirteenth Symposium on Usable Privacy and Security ({SOUPS}), 2017, pp. 211--228.
\newline\urlprefix\url{https://www.usenix.org/conference/soups2017/technical-sessions/presentation/ruoti}

\bibitem{nicholson2019if}
J.~Nicholson, L.~Coventry, P.~Briggs, \href{https://doi.org/10.1145/3290605.3300579}{"if it's important it will be a headline" cyber-security information seeking in older adults}, in: Proceedings of the 2019 {CHI} Conference on Human Factors in Computing Systems, 2019, pp. 1--11, paper No. 349.
\newblock \href {https://doi.org/10.1145/3290605.3300579} {\path{doi:10.1145/3290605.3300579}}.
\newline\urlprefix\url{https://doi.org/10.1145/3290605.3300579}

\bibitem{fulton2019effect}
K.~R. Fulton, R.~Gelles, A.~McKay, R.~Roberts, Y.~Abdi, M.~L. Mazurek, \href{https://www.usenix.org/conference/soups2019/presentation/fulton}{The effect of entertainment media on mental models of computer security}, in: Proceedings of the Fifteenth Symposium on Usable Privacy and Security ({SOUPS}), 2019, pp. 79--95.
\newline\urlprefix\url{https://www.usenix.org/conference/soups2019/presentation/fulton}

\bibitem{lasswell1948structure}
H.~D. Lasswell, The structure and function of communication in society, in: L.~Bryson (Ed.), The Communication of Ideas, Harper and Row, 1948, pp. 37--51.

\bibitem{mccombs1972agenda}
M.~E. McCombs, D.~L. Shaw, \href{https://doi.org/10.1086/267990}{The agenda-setting function of mass media}, Public Opinion Quarterly 36~(2) (1972) 176--187.
\newblock \href {https://doi.org/10.1086/267990} {\path{doi:10.1086/267990}}.
\newline\urlprefix\url{https://doi.org/10.1086/267990}

\bibitem{schirrmacher2018towards}
N.-B. Schirrmacher, J.~Ondrus, F.~T.~C. Tan, Towards a response to ransomware: Examining digital capabilities of the wannacry attack, in: PACIS, 2018, p. 210.

\bibitem{wenger1998communities}
E.~Wenger, \href{https://thesystemsthinker.com/communities-of-practice-learning-as-a-social-system/}{Communities of practice: Learning as a social system}, Systems Thinker 9~(5) (1998) 2--3.
\newline\urlprefix\url{https://thesystemsthinker.com/communities-of-practice-learning-as-a-social-system/}

\bibitem{meyer2010rise}
M.~Meyer, \href{https://doi.org/10.1177/1075547009359797}{The rise of the knowledge broker}, Science Communication 32~(1) (2010) 118--127.
\newblock \href {https://doi.org/10.1177/1075547009359797} {\path{doi:10.1177/1075547009359797}}.
\newline\urlprefix\url{https://doi.org/10.1177/1075547009359797}

\bibitem{contandriopoulos2010knowledge}
D.~Contandriopoulos, M.~Lemire, J.-L. Denis, {\'E}.~Tremblay, \href{https://doi.org/10.1111/j.1468-0009.2010.00608.x}{Knowledge exchange processes in organizations and policy arenas: a narrative systematic review of the literature}, The Milbank Quarterly 88~(4) (2010) 444--483.
\newblock \href {https://doi.org/10.1111/j.1468-0009.2010.00608.x} {\path{doi:10.1111/j.1468-0009.2010.00608.x}}.
\newline\urlprefix\url{https://doi.org/10.1111/j.1468-0009.2010.00608.x}

\bibitem{marres2013scraping}
N.~Marres, E.~Weltevrede, \href{https://doi.org/10.1080/17530350.2013.772070}{Scraping the social? issues in live social research}, Journal of Cultural Economy 6~(3) (2013) 313--335.
\newblock \href {https://doi.org/10.1080/17530350.2013.772070} {\path{doi:10.1080/17530350.2013.772070}}.
\newline\urlprefix\url{https://doi.org/10.1080/17530350.2013.772070}

\bibitem{schatz2017towards}
D.~Schatz, R.~Bashroush, J.~Wall, \href{https://doi.org/10.15394/jdfsl.2017.1476}{Towards a more representative definition of cyber-security}, Journal of Digital Forensics, Security and Law 12~(2) (2017) 53--74.
\newblock \href {https://doi.org/10.15394/jdfsl.2017.1476} {\path{doi:10.15394/jdfsl.2017.1476}}.
\newline\urlprefix\url{https://doi.org/10.15394/jdfsl.2017.1476}

\bibitem{humayun2020cyber}
M.~Humayun, M.~Niazi, N.~Jhanjhi, M.~Alshayeb, S.~Mahmood, Cyber-security threats and vulnerabilities: a systematic mapping study, Arabian Journal for Science and Engineering 45~(4) (2020) 3171--3189.

\bibitem{kosar2016protocol}
T.~Kosar, S.~Bohra, M.~Mernik, Protocol of a systematic mapping study for domain-specific languages, Journal of Information and Software Technology 21~(C) (2016) 77--91.

\bibitem{9006444}
T.~Satyapanich, T.~Finin, F.~Ferraro, \href{https://doi.org/10.1109/BigData47090.2019.9006444}{Extracting rich semantic information about cyber-security events}, in: 2019 IEEE International Conference on Big Data (Big Data), 2019, pp. 5034--5042.
\newblock \href {https://doi.org/10.1109/BigData47090.2019.9006444} {\path{doi:10.1109/BigData47090.2019.9006444}}.
\newline\urlprefix\url{https://doi.org/10.1109/BigData47090.2019.9006444}

\bibitem{Moubayed}
N.~Al~Moubayed, D.~Wall, A.~S. McGough, Identifying changes in the cyber-security threat landscape using the {LDA}-web topic modelling data search engine, in: T.~Tryfonas (Ed.), Human Aspects of Information Security, Privacy and Trust, Springer International Publishing, Cham, 2017, pp. 287--295, lecture Notes in Computer Science, vol 10292.

\bibitem{porter2006algorithm}
M.~F. Porter, An algorithm for suffix stripping, Program (2006).

\bibitem{de2021smart}
A.~De~Nicola, M.~L. Villani, Smart city ontologies and their applications: A systematic literature review, Sustainability 13~(10) (2021) 5578.

\bibitem{ruighaver2007organisational}
A.~B. Ruighaver, S.~B. Maynard, S.~Chang, Organisational security culture: Extending the end-user perspective, Computers \& security 26~(1) (2007) 56--62.

\bibitem{hendrix2016game}
M.~Hendrix, A.~Al-Sherbaz, B.~Victoria, Game based cyber security training: are serious games suitable for cyber security training?, International Journal of Serious Games 3~(1) (2016) 53--61.

\bibitem{oltramari2015towards}
A.~Oltramari, D.~S. Henshel, M.~Cains, B.~Hoffman, Towards a human factors ontology for cyber security., Stids 2015 (2015) 26--33.

\bibitem{grovs2021critical}
S.~Gro{\v{s}}, A critical view on cis controls, in: 2021 16th International Conference on Telecommunications (ConTEL), IEEE, 2021, pp. 122--128.

\bibitem{souag2015security}
A.~Souag, C.~Salinesi, R.~Mazo, I.~Comyn-Wattiau, A security ontology for security requirements elicitation, in: Engineering Secure Software and Systems: 7th International Symposium, ESSoS 2015, Milan, Italy, March 4-6, 2015. Proceedings 7, Springer, 2015, pp. 157--177.

\bibitem{kendall2019ontology}
E.~F. Kendall, D.~L. McGuinness, Ontology engineering, Morgan \& Claypool Publishers, 2019.

\bibitem{piasecki2021defence}
S.~Piasecki, L.~Urquhart, D.~McAuley, Defence against the dark artefacts: Smart home cyber crimes and cyber-security standards, Computer Law \& Security Review 42 (2021) 105542.

\bibitem{adach2022security}
M.~Adach, K.~H{\"a}nninen, K.~Lundqvist, Security ontologies: A systematic literature review, in: International Conference on Enterprise Design, Operations, and Computing, Springer, 2022, pp. 36--53.

\bibitem{woods2017mapping}
D.~Woods, I.~Agrafiotis, J.~R. Nurse, S.~Creese, Mapping the coverage of security controls in cyber insurance proposal forms, Journal of Internet Services and Applications 8 (2017) 1--13.

\bibitem{ruohonen2019updating}
J.~Ruohonen, K.~K. Kimppa, Updating the wassenaar debate once again: Surveillance, intrusion software, and ambiguity, Journal of Information Technology \& Politics 16~(2) (2019) 169--186.

\bibitem{cao2009density}
J.~Cao, T.~Xia, J.~Li, Y.~Zhang, S.~Tang, \href{https://doi.org/10.1016/j.neucom.2008.06.011}{A density-based method for adaptive {LDA} model selection}, Neurocomputing 72~(7--9) (2009) 1775--1781.
\newblock \href {https://doi.org/10.1016/j.neucom.2008.06.011} {\path{doi:10.1016/j.neucom.2008.06.011}}.
\newline\urlprefix\url{https://doi.org/10.1016/j.neucom.2008.06.011}

\bibitem{deveaud2014accurate}
R.~Deveaud, E.~Sanjuan, P.~Bellot, \href{https://doi.org/10.3166/dn.17.1.61-84}{Accurate and effective latent concept modeling for ad hoc information retrieval}, Document Num{\'e}rique 17~(1) (2014) 61--84.
\newblock \href {https://doi.org/10.3166/dn.17.1.61-84} {\path{doi:10.3166/dn.17.1.61-84}}.
\newline\urlprefix\url{https://doi.org/10.3166/dn.17.1.61-84}

\bibitem{alagheband2020time}
M.~R. Alagheband, A.~Mashatan, M.~Zihayat, \href{https://doi.org/10.1145/3389684}{Time-based gap analysis of cyber-security trends in academic and digital media}, ACM Transactions on Management Information Systems (TMIS) 11~(4) (2020) 1--20.
\newblock \href {https://doi.org/10.1145/3389684} {\path{doi:10.1145/3389684}}.
\newline\urlprefix\url{https://doi.org/10.1145/3389684}

\bibitem{taylor2004victorians}
M.~Taylor, M.~Wolff, The Victorians since 1901: Histories, representations and revisions, Manchester University Press, 2004.

\bibitem{goldstein1999summarizing}
J.~Goldstein, M.~Kantrowitz, V.~Mittal, J.~Carbonell, \href{https://doi.org/10.1145/312624.312665}{Summarizing text documents: Sentence selection and evaluation metrics}, in: Proceedings of the 22nd annual international ACM SIGIR conference on Research and development in information retrieval, 1999, pp. 121--128.
\newblock \href {https://doi.org/10.1145/312624.312665} {\path{doi:10.1145/312624.312665}}.
\newline\urlprefix\url{https://doi.org/10.1145/312624.312665}

\bibitem{kalra2019efficacy}
S.~Kalra, J.~S. Prasad, \href{https://doi.org/10.1109/COMITCon.2019.8862265}{Efficacy of news sentiment for stock market prediction}, in: 2019 International Conference on Machine Learning, Big Data, Cloud and Parallel Computing (COMITCon), IEEE, 2019, pp. 491--496.
\newblock \href {https://doi.org/10.1109/COMITCon.2019.8862265} {\path{doi:10.1109/COMITCon.2019.8862265}}.
\newline\urlprefix\url{https://doi.org/10.1109/COMITCon.2019.8862265}

\bibitem{yasaka2020peer}
T.~M. Yasaka, B.~M. Lehrich, R.~Sahyouni, \href{https://doi.org/10.2196/18936}{Peer-to-peer contact tracing: Development of a privacy-preserving smartphone app}, JMIR mHealth and uHealth 8~(4) (2020) e18936.
\newblock \href {https://doi.org/10.2196/18936} {\path{doi:10.2196/18936}}.
\newline\urlprefix\url{https://doi.org/10.2196/18936}

\bibitem{lindner2020tor}
A.~M. Lindner, G.~Pryciak, J.~Elsner, \href{https://ojs.library.queensu.ca/index.php/surveillance-and-society/article/view/13235/9469}{Tor and the city: Msa-level correlates of interest in anonymous web browsing}, Surveillance \& Society 18~(4) (2020) 507--521.
\newline\urlprefix\url{https://ojs.library.queensu.ca/index.php/surveillance-and-society/article/view/13235/9469}

\bibitem{lim2018understanding}
S.~Lim, A.~Jatowt, M.~Yoshikawa, Understanding characteristics of biased sentences in news articles, in: CIKM Workshops, 2018, pp. 121--128.

\bibitem{heaps1978information}
H.~S. Heaps, Information retrieval, computational and theoretical aspects, Academic Press, 1978.

\bibitem{kanungo2009predicting}
T.~Kanungo, D.~Orr, \href{https://doi.org/10.1145/1498759.1498827}{Predicting the readability of short web summaries}, in: Proceedings of the Second ACM International Conference on Web Search and Data Mining, 2009, pp. 202--211.
\newblock \href {https://doi.org/10.1145/1498759.1498827} {\path{doi:10.1145/1498759.1498827}}.
\newline\urlprefix\url{https://doi.org/10.1145/1498759.1498827}

\bibitem{grinberg2018identifying}
N.~Grinberg, \href{https://doi.org/10.1145/3178876.3186180}{Identifying modes of user engagement with online news and their relationship to information gain in text}, in: Proceedings of the 2018 World Wide Web Conference, 2018, pp. 1745--1754.
\newblock \href {https://doi.org/10.1145/3178876.3186180} {\path{doi:10.1145/3178876.3186180}}.
\newline\urlprefix\url{https://doi.org/10.1145/3178876.3186180}

\bibitem{flesch2007flesch}
R.~Flesch, Flesch-kincaid readability test, Retrieved October 26~(3) (2007) 2007.

\bibitem{roberts1994effects}
J.~C. Roberts, R.~H. Fletcher, S.~W. Fletcher, \href{https://doi.org/10.1001/jama.1994.03520020045012}{Effects of peer review and editing on the readability of articles published in annals of internal medicine}, JAMA 272~(2) (1994) 119--121.
\newblock \href {https://doi.org/10.1001/jama.1994.03520020045012} {\path{doi:10.1001/jama.1994.03520020045012}}.
\newline\urlprefix\url{https://doi.org/10.1001/jama.1994.03520020045012}

\bibitem{coleman1975computer}
M.~Coleman, T.~L. Liau, \href{https://doi.org/10.1037/h0076540}{A computer readability formula designed for machine scoring}, Journal of Applied Psychology 60~(2) (1975) 283--284.
\newblock \href {https://doi.org/10.1037/h0076540} {\path{doi:10.1037/h0076540}}.
\newline\urlprefix\url{https://doi.org/10.1037/h0076540}

\bibitem{senter1967automated}
R.~J. Senter, E.~A. Smith, \href{https://apps.dtic.mil/dtic/tr/fulltext/u2/667273.pdf}{Automated readability index}, Tech. rep., AMRL-TR. Aerospace Medical Research Laboratories (1967).
\newline\urlprefix\url{https://apps.dtic.mil/dtic/tr/fulltext/u2/667273.pdf}

\bibitem{mc1969smog}
G.~H. Mc~Laughlin, \href{https://www.jstor.org/stable/40011226}{Smog grading---a new readability formula}, Journal of Reading 12~(8) (1969) 639--646.
\newline\urlprefix\url{https://www.jstor.org/stable/40011226}

\bibitem{hussein2018survey}
D.~M. E.-D.~M. Hussein, \href{https://doi.org/10.1016/j.jksues.2016.04.002}{A survey on sentiment analysis challenges}, Journal of King Saud University - Engineering Sciences 30~(4) (2018) 330--338.
\newblock \href {https://doi.org/10.1016/j.jksues.2016.04.002} {\path{doi:10.1016/j.jksues.2016.04.002}}.
\newline\urlprefix\url{https://doi.org/10.1016/j.jksues.2016.04.002}

\bibitem{macdonald2015identifying}
M.~Macdonald, R.~Frank, J.~Mei, B.~Monk, \href{https://doi.org/10.1145/2808797.2808878}{Identifying digital threats in a hacker web forum}, in: Proceedings of the 2015 IEEE/ACM International Conference on Advances in Social Networks Analysis and Mining 2015, 2015, pp. 926--933.
\newblock \href {https://doi.org/10.1145/2808797.2808878} {\path{doi:10.1145/2808797.2808878}}.
\newline\urlprefix\url{https://doi.org/10.1145/2808797.2808878}

\bibitem{deerwester1990indexing}
S.~Deerwester, S.~T. Dumais, G.~W. Furnas, T.~K. Landauer, R.~Harshman, Indexing by latent semantic analysis, Journal of the American society for information science 41~(6) (1990) 391--407.

\bibitem{hamilton2016inducing}
W.~L. Hamilton, K.~Clark, J.~Leskovec, D.~Jurafsky, \href{https://www.aclweb.org/anthology/D16-1057.pdf}{Inducing domain-specific sentiment lexicons from unlabeled corpora}, in: Proceedings of the 2016 Conference on Empirical Methods in Natural Language Processing, Vol. 2016, NIH Public Access, 2016, pp. 595--605.
\newline\urlprefix\url{https://www.aclweb.org/anthology/D16-1057.pdf}

\bibitem{oldehoeft1992foundations}
A.~E. Oldehoeft, \href{https://nvlpubs.nist.gov/nistpubs/Legacy/IR/nistir4734.pdf}{Foundations of a Security Policy for Use of the National Research and Educational Network}, U.S. Department of Commerce, National Institute of Standards and Technology, 1992.
\newline\urlprefix\url{https://nvlpubs.nist.gov/nistpubs/Legacy/IR/nistir4734.pdf}

\bibitem{cook1998governing}
T.~E. Cook, Governing with the news: The news media as a political institution, University of Chicago Press, 1998.

\bibitem{gadarian2010politics}
S.~K. Gadarian, \href{https://doi.org/10.1017/s0022381609990910}{The politics of threat: How terrorism news shapes foreign policy attitudes}, The Journal of Politics 72~(2) (2010) 469--483.
\newblock \href {https://doi.org/10.1017/s0022381609990910} {\path{doi:10.1017/s0022381609990910}}.
\newline\urlprefix\url{https://doi.org/10.1017/s0022381609990910}

\bibitem{kher2017readability}
A.~Kher, S.~Johnson, R.~Griffith, \href{https://doi.org/10.1155/2017/9780317}{Readability assessment of online patient education material on congestive heart failure}, Advances in Preventive Medicine 2017 (2017).
\newblock \href {https://doi.org/10.1155/2017/9780317} {\path{doi:10.1155/2017/9780317}}.
\newline\urlprefix\url{https://doi.org/10.1155/2017/9780317}

\bibitem{britt2017ehealth}
R.~K. Britt, W.~B. Collins, K.~Wilson, G.~Linnemeier, A.~M. Englebert, \href{https://doi.org/10.2196/preprints.3100}{ehealth literacy and health behaviors affecting modern college students: A pilot study of issues identified by the american college health association}, Journal of medical Internet research 19~(12) (2017) e392.
\newline\urlprefix\url{https://doi.org/10.2196/preprints.3100}

\bibitem{Sasse}
A.~Adams, M.~A. Sasse, \href{https://doi.org/10.1145/322796.322806}{Users are not the enemy}, Communications of the ACM 42~(12) (1999) 40--46.
\newblock \href {https://doi.org/10.1145/322796.322806} {\path{doi:10.1145/322796.322806}}.
\newline\urlprefix\url{https://doi.org/10.1145/322796.322806}

\bibitem{hochstotter2009users}
N.~H{\"o}chst{\"o}tter, D.~Lewandowski, What users see--structures in search engine results pages, Information Sciences 179~(12) (2009) 1796--1812.

\end{thebibliography}

\appendix
\section{Search terms}

\renewcommand{\arraystretch}{1.3} 
\setlength{\tabcolsep}{5pt} 

\begin{longtable}{|p{2.5cm}|p{9.5cm}|}
\caption{Refined inclusion and exclusion criteria for cybersecurity search terms using AND/OR logic.}\label{tab:logicsearchcriteria} \\

\hline
\textbf{Criteria} & \textbf{Search Terms} \\
\hline
\endfirsthead

\hline
\textbf{Criteria} & \textbf{Search Terms} \\
\hline
\endhead

\hline
\endfoot

\hline
\endlastfoot

\textbf{Incl.} & 
\begin{itemize}
    \item (Cybersecurity OR "Cyber Security" OR "Cyber Safety") AND (Tips OR Advice OR Best Practices OR Guidelines OR Recommendations)
    \item ("Online Protection" OR "Internet Security") AND (Individuals OR Families OR Home Users)
    \item ("Hacking Prevention" OR "Anti-Hacking" OR "Hack Prevention") AND (Personal OR Individual)
    \item ("Password Security" OR "Strong Passwords" OR Authentication) AND (Tips OR Management OR Best Practices)
    \item ("Social Network Security" OR "Social Media Protection") AND (Guide OR How-to OR Instructions)
    \item ("Email Security" OR "Phishing Prevention") AND (Awareness OR Training OR Education)
    \item (Malware OR "Anti-Malware Software" OR Antivirus) AND (Recommendations OR Reviews OR Comparisons)
    \item ("Cyber Hygiene" OR "Cyber Awareness") AND (Promoting OR Improving OR Increasing)
    \item (Firewall OR "Intrusion Detection" OR "Intrusion Prevention") AND (Home Networks OR Personal Devices)
\end{itemize} \\
\hline

\textbf{Excl.} &
\begin{itemize}
    \item (Cyberattack OR "Cyber Attack" OR "Security Breach") AND (Nation-state OR APT OR "Advanced Persistent Threat")
    \item ("Cyber Espionage" OR "Political Hacking" OR Hacktivism)
    \item (Cyberwar OR "Cyber Warfare" OR "Nation-State Hacking")
    \item ("Corporate Cybersecurity" OR "Enterprise Security")
    \item ("Critical Infrastructure" OR Military OR Government) AND (Cybersecurity OR Protection OR Defense)
    \item ("Offensive Cyber" OR "Hacking Tools") AND (Capabilities OR Operations OR Techniques)
    \item (Cybercrime OR "Cyber Terrorism") AND (Trends OR Statistics OR Incidents)
\end{itemize} \\
\hline

\end{longtable}

\renewcommand{\arraystretch}{1.3} 
\setlength{\tabcolsep}{5pt} 

\begin{longtable}{|p{1.5cm}|p{2cm}|p{4cm}|p{4cm}|}
\caption{A table of 20 cybersecurity events that took place in the 24 months leading up to the article being written.}
\label{tab:newsqueries} \\
\hline
\textbf{Event ID} & \textbf{Event Name} & \textbf{Event Description} & \textbf{Search Terms} \\
\hline
\endfirsthead
\caption[]{(continued)} \\
\hline
\textbf{Event ID} & \textbf{Event Name} & \textbf{Event Description} & \textbf{Search Terms} \\
\hline
\endhead
\hline
\endfoot
\hline
\endlastfoot
1 & Airbus Cyber Attacks & Airbus was hit by a series of cyber attacks targeting its suppliers to steal technical documents. & Airbus cyber attack 2019, Airbus suppliers hack, Airbus data breach \\
\hline
2 & BlueKeep Windows Vulnerability & A critical remote code execution vulnerability was discovered in Windows, allowing attackers to take control of systems without any user interaction. & BlueKeep vulnerability 2019, CVE-2019-0708, Windows remote desktop flaw \\
\hline
3 & U.S. Customs and Border Protection Data Breach & U.S. Customs and Border Protection suffered a data breach that exposed photos of people and vehicles traveling into and out of the country. & U.S. Customs data breach 2019, CBP photo hack, border protection cyber attack \\
\hline
4 & Kr00k Wi-Fi Encryption Vulnerability & A vulnerability was found in billions of Wi-Fi devices that could allow attackers to decrypt wireless network packets. & Kr00k vulnerability 2020, CVE-2019-15126, Wi-Fi KRACK attack \\
\hline
5 & Texas Local Governments Ransomware Attack & Over 20 Texas local governments were targeted in a coordinated ransomware attack. & Texas ransomware attack 2019, Texas local government hack, coordinated ransomware \\
\hline
6 & WhatsApp Security Flaw CVE-2019-3566 & A buffer overflow vulnerability in WhatsApp allowed remote code execution by attackers simply by calling the victim's phone. & WhatsApp CVE-2019-3566, WhatsApp remote code execution, WhatsApp buffer overflow \\
\hline
7 & Nunavut Government Ransomware Attack & Ransomware hackers attacked the government of Nunavut, Canada, crippling its computer systems. & Nunavut ransomware 2019, Nunavut government hack, Canadian government cyber attack \\
\hline
8 & Travelex Ransomware Attack & Travelex, a foreign currency exchange company, was hit by ransomware causing its services to be taken offline for weeks. & Travelex ransomware 2019, Travelex hack, foreign exchange cyber attack \\
\hline
9 & United Nations Office Hacked & The United Nations was hacked via its UN Office at Geneva and UN Office at Vienna, with hackers gaining access to staff records, health insurance, and commercial contract data. & UN Geneva hack 2020, UN Vienna data breach, United Nations cyber attack \\
\hline
10 & U.S. Department of Defense Data Breach & The U.S. Department of Defense agency that handles secure communications for the White House suffered a data breach. & U.S. Department of Defense data breach 2020, White House communications hack, DoD cyber attack \\
\hline
11 & Marriott Data Breach & Marriott International announced a data breach exposing the personal information of 5.2 million guests. & Marriott data breach 2020, Marriott hack, hotel data breach \\
\hline
12 & Cognizant Ransomware Attack & Cognizant, one of the largest IT managed services company, was hit by the Maze ransomware. & Cognizant ransomware 2020, Maze ransomware attack, IT services hack \\
\hline
13 & EasyJet Data Breach & EasyJet announced 9 million customers' email addresses and travel details had been breached. & EasyJet data breach 2020, EasyJet hack, airline data breach \\
\hline
14 & Honda Ransomware Attack & Honda was forced to suspend some production after being hit by a ransomware attack. & Honda ransomware 2020, Honda production hack, automotive cyber attack \\
\hline
15 & Garmin Ransomware Attack & Garmin, the GPS and fitness-tracker company, was hit by a ransomware attack that disrupted its services for days. & Garmin ransomware 2020, Garmin hack, GPS company cyber attack \\
\hline
16 & Canon Ransomware Attack & Canon suffered a ransomware attack that resulted in 10TB of data being stolen. & Canon ransomware 2020, Canon data breach, camera company hack \\
\hline
17 & German Hospital Ransomware Attack & A ransomware attack hit a German hospital, causing IT systems to fail and a woman to die when she had to be taken to another city for treatment. & German hospital ransomware 2020, hospital IT failure hack, medical cyber attack \\
\hline
18 & Ripple20 Vulnerabilities & 19 zero-day vulnerabilities were discovered in a widely used low-level TCP/IP software library, impacting millions of IoT devices. & Ripple20 vulnerabilities 2020, Treck TCP/IP library flaws, IoT device vulnerabilities \\
\hline
19 & ZombieLoad, Fallout, RIDL Intel CPU Flaws & New Intel CPU vulnerabilities were disclosed that could allow attackers to steal sensitive data. The flaws are similar to Spectre and Meltdown. & ZombieLoad vulnerability, Fallout Intel flaw, MDS attack, Intel CPU data leak \\
\hline
20 & FireEye Data Breach & U.S. cybersecurity firm FireEye disclosed that it was hacked, likely by a nation-state, and had its own hacking tools stolen. & FireEye data breach 2020, cybersecurity firm hack, FireEye hacking tools stolen \\
\hline
\end{longtable}





\end{document}